\documentclass[a4paper,fleqn,usenatbib]{mnras}

\usepackage{mathptmx}

\usepackage[T1]{fontenc}
\usepackage{ae,aecompl}

\usepackage{graphicx}	
\usepackage{amsmath}	
\usepackage{amssymb}	

\title[Observability of warps in circumbinary discs]{Observational signatures of linear warps in circumbinary discs}

\author[Juh\'{a}sz, Facchini]{Attila Juh\'{a}sz$^1$\thanks{juhasz@ast.cam.ac.uk}, Stefano Facchini$^{2}$\thanks{facchini@mpe.mpg.de}\\
$^1$Institute of Astronomy, Madingley Road, Cambridge CB3 OHA, UK\\
$^2$Max-Planck-Institut f\"ur Extraterrestrische Physik, Giessenbachstrasse 1, 85748 Garching, Germany\\
}

\date{Accepted XXX. Received YYY; in original form ZZZ}

\pubyear{2016}

\begin{document}
\label{firstpage}
\pagerange{\pageref{firstpage}--\pageref{lastpage}}
\maketitle

\begin{abstract}
In recent years an increasing number of observational studies have hinted at the presence of warps in protoplanetary discs, however a general 
comprehensive description of observational diagnostics of warped discs was missing. We performed a series of 3D SPH hydrodynamic simulations
and combined them with 3D radiative transfer calculations to study the observability of warps in circumbinary discs, whose plane is misaligned with respect to the orbital plane of the central binary. Our numerical hydrodynamic simulations confirm previous analytical results on the dependence of the warp structure on the viscosity and the initial misalignment between the binary and the disc. To study the observational signatures of warps we calculate images in the continuum at near-infrared and sub-millimetre wavelengths and in the pure rotational transition of CO in the sub-millimetre. Warped circumbinary discs show surface brightness asymmetry in near-infrared scattered light images as well as in optically thick gas lines at sub-millimetre wavelengths. The asymmetry is caused by self-shadowing of the disc by the inner warped regions, thus the strength of the asymmetry depends on the strength of the warp.  The projected velocity field, derived from line observations, shows characteristic deviations, twists and a change in the slope of the rotation curve, from that of an unperturbed disc. In extreme cases even the direction of rotation appears to change in the disc inwards of a characteristic radius. The strength of the kinematical signatures of warps decreases with increasing inclination. The strength of all warp signatures
decreases with decreasing viscosity.
\end{abstract}




\section{Introduction}
\label{sec:introduction}

Protoplanetary discs are usually thought of as axisymmetric structures orbiting a single star. However, it is well known that the geometry of such discs can be more complicated. In particular, discs in binary systems can be warped. The orbital plane of the binary and the plane of the outer disc are likely to be misaligned due to interaction with other stars or to accretion of randomly oriented gas \citep[e.g.][]{1992ApJ...400..579B,2010MNRAS.401.1505B}. Such misalignment induces a torque in the disc, leading to a warped 3D structure.

Until a few years ago, we lacked observational evidence of warped protoplanetary discs. In binary systems, tilted discs had been inferred only around X-ray binaries showing a long term modulation of the periodic light curve \citep[see][and references therein]{2012MNRAS.420.1575K}, and some cataclysmic variables \citep[summarised by][]{2009MNRAS.399..465O}. However, very recently, observational features in a few protoplanetary discs have been interpreted as signatures of disc warping. In particular, the advent of interferometric observations with the Acatama Large Millimeter Array (ALMA) has allowed us to start measuring precise gas kinematics via molecular line profiles, which is one of the most powerful tools to detect disc warping. For example, \citet{2012ApJ...757..129R} have observed an excess in the emission at high velocity in the inner disc of TW Hya, which they interpreted as a possible inner warp. Another case is HD 100546, where a warp  has been invoked to interpret the twisted first moment map of CO (3-2) \citep{2014ApJ...788L..34P}, and even earlier as a possible explanation of the spiral structure observed in scattered light \citep{2006ApJ...640.1078Q} and of the asymmetric line profiles from single-dish observations \citep{2010A&A...519A.110P}. A very significant case is HD 142527. \citet{2015ApJ...798L..44M} have modelled the strongly asymmetric scattered light image of such a protoplanetary disc with a highly misaligned inner disc, casting a shadow onto the outer regions. The misaligned inner disc could be tracing the inclination of the stellar companion \citep{2012ApJ...753L..38B,2014ApJ...781L..30C} observed within the huge mm cavity of the transition disc. Furthermore, CO (6-5) ALMA observations of the system have shown a twisted velocity centroid map \citep{2015ApJ...811...92C}, which can be modelled by extreme disc warping, probably caused by a dynamical interaction with the low mass companion \citep{2016PASA...33...13C}. Non-Keplerian flows have also been detected with lower angular resolution in CO (3-2) \citep{2013Natur.493..191C,2014ApJ...782...62R}. Finally, the KH 15D binary system has been modelled as a compact precessing misaligned circumbinary disc, which causes the long-term modulation of the light curve of the system \citep{2004ApJ...607..913C,2010AJ....140.2025H,2013MNRAS.433.2157L,2014AJ....147....9W}. All these observations go together with complementary recent findings, where disc warping has not been detected yet, but where circumprimary (and sometimes circumsecondary) discs are clearly misaligned with the orbital plane of the binary \citep[e.g.][]{2011A&A...534A..33R,2014Natur.511..567J,2014ApJ...796..120W}. Some of these discs are very likely to show kinematic signatures of disc warping, when higher angular resolution observations will be available. Finally, the occurrence of particularly deep dimming events in the optical of some disc-bearing systems (as AA Tau and RW Aur) has been also interpreted as due to warping of the inner disc ($<$ few au), where the warp could be caused both by magnetic fields \citep[see][for AA Tau]{2013A&A...557A..77B} or unresolved binaries \citep[see][for RW Aur]{2016A&A...596A..38F}.

The theory of disc warping has anticipated the observations by several decades (in the protoplanetary regime). For a recent review on the theoretical basis of the phenomenon, see \citet{2016LNP...905...45N}. The response of a disc to a non axisymmetric (effective) potential has been initially studied by \citet{1983MNRAS.202.1181P}. This study derived the equations for the evolution of a warp in a diffusive regime, which occurs whenever viscous forces dominate over pressure forces in the disc. Here we define the aspect ratio of a disc as $H_{\rm p}/r$, where $H_{\rm p}$ is the scale height of the disc and $r$ the radial coordinate, and we parametrise the disc kinematic viscosity with the simple prescription by \citet{1973A&A....24..337S}, i.e. $\nu=\alpha c_{\rm s}H_{\rm p}$, where $c_{\rm s}$ is the sound speed, and $\alpha$ a dimensionless parameter $\ll1$. The criterion for a diffusively evolving disc becomes then $\alpha>H_{\rm p}/r$ \citep{1983MNRAS.202.1181P}. \citet{1999MNRAS.304..557O} has extended the analytic approach to a full non-linear regime in the diffusive case. When $\alpha<H_{\rm p}/r$, the warp propagates as a bending wave with a velocity $\sim c_{\rm s}/2$ \citep[e.g.][]{papaloizou_1995, 1999ASPC..160...53P}, since pressure forces dominate the communication of the external torque.  Linearised equations modelling the warp evolution in such regime have been derived by \citet{papaloizou_1995}, \citet{1997A&A...324..829D} and \citet{2000ApJ...538..326L}. An initial study on the non-linear response of these bending waves is reported in \citet{2006MNRAS.365..977O}. For typical physical parameters, the diffusive regime describes accretion discs around black holes, whereas the bending waves regime is applicable to protoplanetary and protostellar discs. Many of these analytic results have been confirmed via numerical simulations. In particular, in a linear regime, 3D hydrodynamical simulations have shown that the warp evolution equations (in both diffusive and bending wave regime) well describe the evolution of the disc on short timescales . On longer (close to viscous) timescales, it is preferable to model the warp evolution with 3D hydrodynamical simulations, since they include all the corrected physics in a simple $\alpha$-disc \citep{2016LNP...905...45N}. A more detailed discussion on such 3D simulations is reported in Section~\ref{sec:sph}. 

Whereas new observations are probing possible signatures of disc warping, and the theoretical basis of such mechanism is well established (at least for $\alpha$-discs), the connection between the two has been poorly addressed. In particular, not much attention has been drawn to observational diagnostics of warped protoplanetary discs, in order to extract quantitative information when such structures are observed. The effect of the asymmetric illumination of a warped disc was studied by \citet{terquem_1996} in circumbinary discs while  \citet{2010MNRAS.403.1887N} studied the effect of irradiation in discs warped due to tidal interactions during a stellar fly-by. These studies noticed that warping induces significant changes in the spectral energy distribution (SED) at wavelengths $\lambda>100\,\mu$m. Similarly, \citet{2010ApJ...719.1733F} provided a prediction for the SEDs of warped discs, where the warping was not computed via hydrodynamical simulations, but was simply parametrised in a razor-thin model. \citet{2014ApJ...782...62R} \citep[see also][]{2015ApJ...811...92C} predicted that warps would cause typical twisted first moment maps in the molecular line emission. Finally, \cite{2015A&A...579A.110R} post-processed 3D hydrodynamical simulations of both coplanar and warped circumbinary discs, showing how the latter cause an asymmetric illumination that can be traced in continuum intensities maps by ALMA. 

In this paper, we aim to systematically derive observational diagnostics of warped circumbinary discs. The warped structures are estimated using 3D smoothed particle hydrodynamics (SPH) simulations, and then post processed to obtain both total intensity maps at submm wavelengths, and radially resolved molecular emission lines. We also provide similar diagnostics in total intensity maps in scattered light. We cover a large parameter space, ranging over viscosity, inclinations and position angles. These predictions can be used to interpret present and future observations.

\section{Hydrodynamical simulations}
\label{sec:sph}

In order to obtain the warped structure of a misaligned circumbinary disc, we use the 3D SPH code \textsc{phantom}\footnote{\url{http://users.monash.edu.au/~dprice/phantom}} \citep{2010MNRAS.406.1659P,2010MNRAS.405.1212L,price_2012}, which has shown significant agreement with both linear and non linear theory of warp propagation \citep{2010MNRAS.405.1212L,2013MNRAS.433.2142F,2015MNRAS.448.1526N}, and has already been used to model circumbinary discs \citep[][]{2012MNRAS.423.2597N,2013MNRAS.434.1946N,2013MNRAS.433.2142F,2016arXiv161008381C,2017MNRAS.464.1449R}. 

\subsection{Setup}

For the simulations, we use a setup very similar to the one reported in \citet{2013MNRAS.433.2142F}. All the simulations are initiated with two equal mass stars on a circular orbit with a binary separation $a$. For a coplanar circumbinary disc, the tidal truncation radius is $\sim1.7a$ \citep{1994ApJ...421..651A}, and it is not expected to shrink significantly for moderate misalignments \citep{2015ApJ...800...96L}. We therefore model the stars with sink particles \citep[e.g.][]{1995MNRAS.277..362B}, with an accretion radius of $0.95a$, and a total binary mass of unity in code units. We are not interested in following the gaseous streams flowing onto the binary stars, and we choose such large accretion radii to speed up the computation. The initial surface density of the disc is:

\begin{equation}
\Sigma(r)=\Sigma_0 r^{-1}\left( 1-\sqrt{\frac{r_{\rm in}}{r}} \right),
\end{equation}
where $r_{\rm in}=1.5a$ and $\Sigma_0$ is an arbitrary scale parameter in the hydrodynamical simulations (we neglect self gravity). We set an initial outer radius $r_{\rm out}=15a$. The disc is assigned a locally isothermal equation of state, with $c_{\rm s} \propto r^{-1/4}$, in order to have a temperature profile that scales as $\sim r^{-1/2}$. Temperature and surface density profiles were chosen to match typical fitted profiles of protoplanetary discs \citep[e.g.][]{2005ApJ...631.1134A,2007ApJ...671.1800A}. The aspect ratio $H_{\rm p}/r$ is set to $0.1$ at $r=r_{\rm in}$. All the discs are simulated with $N=10^6$ SPH particles.

Some simulations \citep[e.g.][]{2013MNRAS.434.1946N,2013MNRAS.433.2142F} have shown that a circumbinary disc can tear apart \citep[the same behaviour has been observed both with other codes and in different contexts, such as circumprimary discs, accretion discs around spinning black holes, see e.g.][]{1996MNRAS.282..597L,1997MNRAS.285..288L,2010A&A...511A..77F,2010MNRAS.405.1212L,2012ApJ...757L..24N,2015MNRAS.449.1251D,2015MNRAS.448.1526N}, when the local precession torque exceeds the local viscous torque (in a diffusive regime). In the bending wave regime, discs seem to tear apart when the local precession period induced by the external torque is longer than the wave communication timescale \citep[see the Appendix in][]{2013MNRAS.434.1946N,2015MNRAS.448.1526N}. Even though the details of tearing in circumbinary discs are yet to be thoroughly investigated
in the bending wave regime, we can confidently prevent the disc from tearing apart, by simulating small/moderate misalignment angles between the binary orbit and the plane of the disc.
We run simulations with two different initial misalignment angles, denoted by $\beta_0$, of 15\degr and 30\degr. We have checked {\it a posteriori} that none of our simulations shows any sign of disc breaking.

In order to precisely control the physical viscosity in our simulations we use the setup described in section 6.1 of \citet{2013MNRAS.433.2142F}. We compute directly the stress tensor in the Navier-Stokes equation, following the formulation by \citet{1994ApJ...431..754F}, which has proved to mimic the physical viscosity with high precision \citep{2010MNRAS.405.1212L}. We set the shear viscosity by using the standard $\alpha$ parameter, and we set the bulk component to $0$. The kinematic viscosity $\nu$ is computed via the equation:

\begin{equation}
\nu=\alpha \frac{c_{\rm s}^2}{\Omega}
\end{equation}
In the simulations we still have an artificial viscosity, in order to deal with potential shocks and prevent particle interpenetration using the \citet{1997JCoPh.136...41M} switch. In order for the physical viscosity to be higher than the effective term generated by the artificial viscosity \citep{2010MNRAS.405.1212L} our simulations have quite high $\alpha$ values.

As mentioned above, the disc is initially flat on a plane misaligned to the binary orbit. The disc then evolves, responding to the external torque of the central binary, until it reaches a warped quasi steady state. The simulations are run long enough (for $\sim600$ binary orbits for most cases, $\sim1130$ for the less viscous simulations) that both the surface density and the warp reach such quasi steady state. Thus the warped 
structure of the disc is maintained until the end of our simulation.
In  \autoref{tab:hydro} we summarise the main parameters of the SPH simulations. We did not explore a large parameter space here, since we preferred doing so in the post-processing phase for the azimuthal angles and the inclinations along the line of sight of the simulated discs.
 
\begin{figure}
\center
\includegraphics[width=8cm]{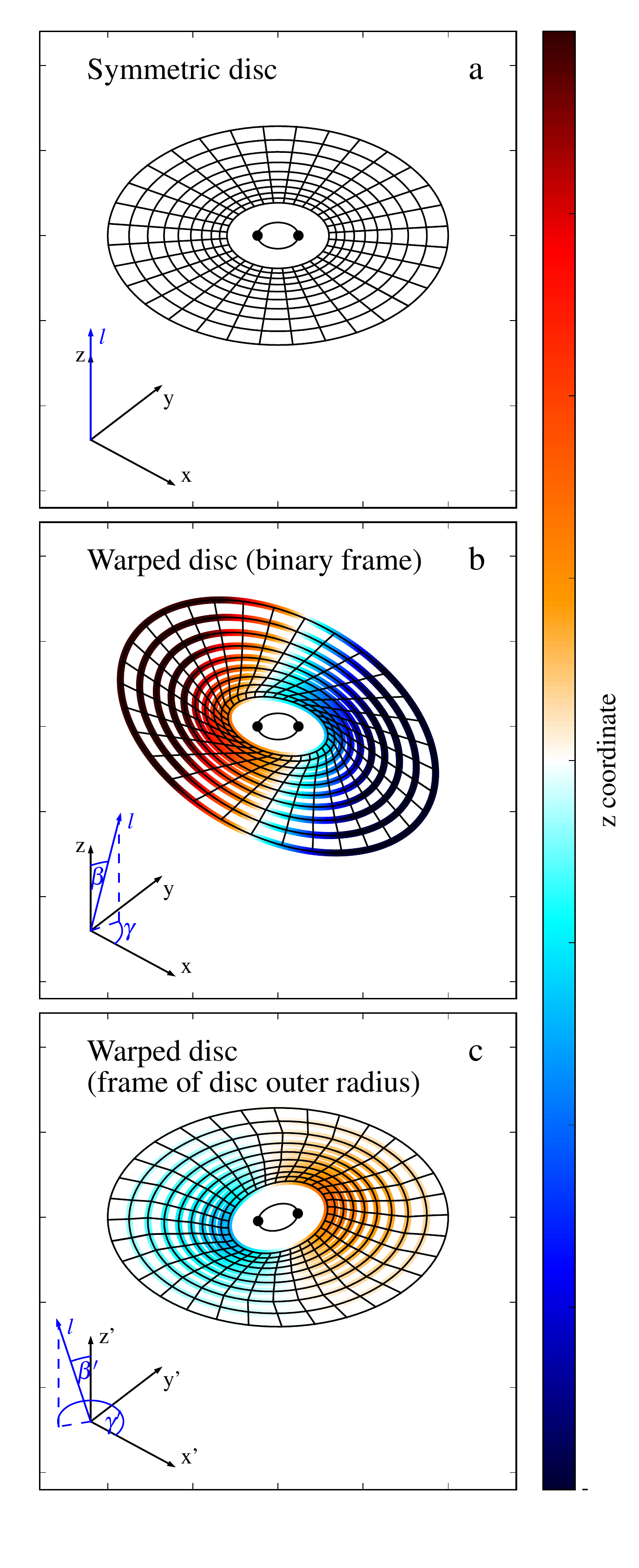}
\caption{Illustration of the warp structure discussed in the paper. The bottom left cartoon shows the orientation of the coordinate system while the blue 
arrow shows the specific angular momentum vector at the inner edge of the disc.  {\it Panel a:} The structure of an unperturbed disc whose rotation axis is aligned with the z-axis of a cartesian coordinate system is axisymmetric around the z-axis and has a mirror symmetry with respect to the x-y plane. {\it Panel b:} A warped disc is characterised by two angles: the twist ($\beta(r)$) and the tilt ($\gamma(r)$) defined in a coordinate system aligned with the orbital plane of the central binary. The tilt angle is the angle between the positive x-axis and the projection of the specific angular momentum vector onto the x-y plane. {\it Panel c:} Without the knowledge of the binary orbit one can only determine quantities defined in a coordinate system aligned with the disc itself at a characteristic radius. In this figure the coordinate system is aligned with the outer edge of the disc.}
\label{fig:cartoon}
\end{figure}

\begin{figure}
\center
\includegraphics[width=\columnwidth]{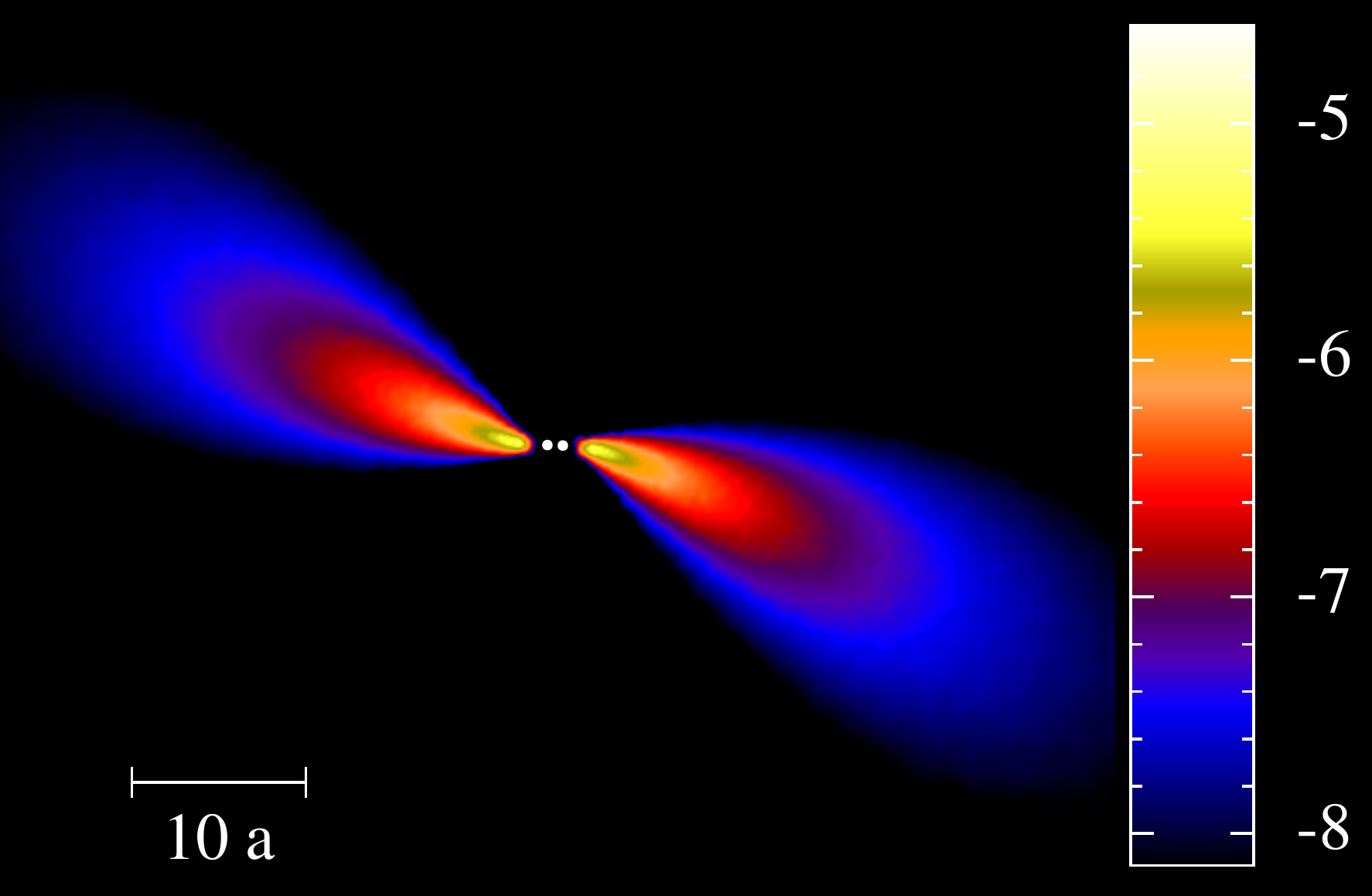}
\caption{Cross section of the final snapshot of the most viscous disc ($\alpha=0.2$). Colour coding shows density on a logarithmic scale, in arbitrary units.}
\label{fig:sph}
\end{figure}

\begin{table}
\caption{Setup of the SPH simulations. All the simulations are run for an equal mass binary with separation $a$, $r_{\rm out}=15a$, and $N=10^6$ particles.}
\centering
\begin{tabular}{llll}
\hline
$\alpha$ & $\beta_0$ (\degr) & Binary orbits & $N$ \\
\hline
$0.02$ & $15$ & 1130 & $10^6$ \\
$0.02$ & $30$ & 1130 & $10^6$ \\
$0.05$ & $15$ & $600$ & $10^6$ \\
$0.05$ & $30$ & $600$ & $10^6$ \\
$0.1$ & $15$ & $600$ & $10^6$ \\
$0.1$ & $30$ & $600$ & $10^6$ \\
$0.2$ & $15$ & $600$ & $10^6$ \\
$0.2$ & $30$ & $600$ & $10^6$ \\
\hline
\hline
\label{tab:hydro}
\end{tabular}
\end{table} 

\subsection{Results}
\label{subsec:hydro_results}

A warped disc can be parametrised by two angles, which are function of the (spherical) radius $r$ (see \autoref{fig:cartoon}). In the thin disc approximation, we can assume that the disc is composed by a series of flat, infinitesimally thin rings, each of which can be oriented arbitrarily in space. The orientation of each ring can be described by its specific angular momentum ${\bf l}(r)$. To describe the orientation of ${\bf l}(r)$ in the 3D cartesian space we chose a frame of reference in which the angular momentum vector of the central binary is aligned with the $z$-axis. In this case the specific angular momentum of each ring can be written in the following form  \citep[e.g.][]{1996MNRAS.281..357P}: ${\bf l}(r) = (\cos\gamma(r)\sin\beta(r),\sin\gamma(r)\sin\beta(r),\cos\beta(r))^{\rm T}$. 
The tilt angle, $\beta(r)$, defines the angle between the direction of the specific angular momentum of the disc and the binary angular momentum
(i.e. the positive $z$-axis), while the twist, $\gamma(r)$, describes the azimuthal angle of the specific angular momentum with respect to an arbitrary 
axis perpendicular to the angular momentum of the binary (i.e. in the $xy$-plane).  Thus, the misalignment between the binary orbit and a planar disc, as set in our initial conditions of the SPH simulations, can be described as a radially independent non-zero tilt angle $\beta_0$. In order to obtain $\beta(r)$, $\gamma(r)$ as well as the radial surface density profile $\Sigma(r)$ from the SPH simulations we compute azimuthally
averaged disc quantities in thin spherical shells. The procedure is described in section 3.2.6 of \citet{2010MNRAS.405.1212L}.

\begin{figure*}
\begin{center}
\includegraphics[width=\textwidth]{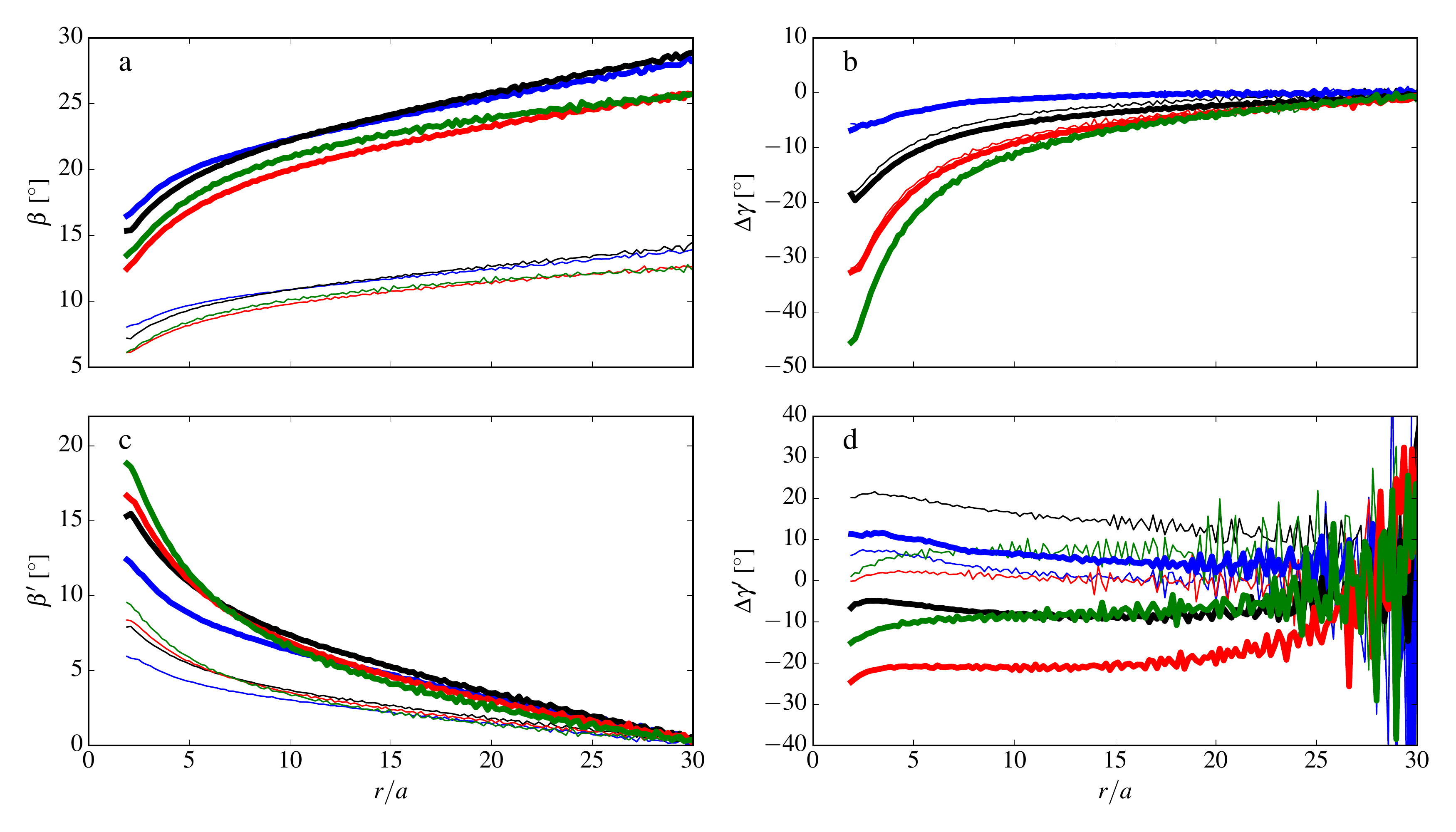}
\end{center}
\caption{Tilt and twist angles of the SPH simulations. Colours indicate different viscosities: blue, black, red and green lines are associated with $
\alpha=0.02$, $0.05$, $0.1$ and $0.2$, respectively. Thick lines show the simulations with initial inclination $i=30^\circ$, and thin lines simulations with 
$i=15^\circ$. Panel {\it a} and {\it b} shows the tilt ($\beta$) and the variation of twist between the inner and the outer radius ($\Delta \gamma$) defined in the 
coordinate system aligned to the orbit of the binary.
It is easily seen that for a fixed binary parameters and orbit $\beta$ depends very weakly on the viscosity but instead is determined by the initial misalignment 
angle between the disc and the binary. In contrast, $\Delta\gamma$, the total variation of $\gamma$ in the disc, is set by the viscosity and is independent 
of the initial misalignment if the binary orbit and mass ratio is fixed. Panel {\it c} and {\it d} shows the tilt ($\beta^\prime$) and the variation of the twist 
($\Delta\gamma^\prime$) angles defined in a coordinate system aligned to the plane of the outer edge of the disc (see Section~\ref{sec:prime_angles} and \autoref{fig:cartoon} for definition). While the most important parameter determining $\beta^\prime$ is still the initial misalignment the scatter is larger than for 
$\beta$ in Panel {\it a}. $\Delta\gamma^\prime$ seems to depend extremely weakly on the viscosity.
}
\label{fig:tilt}
\end{figure*}

In \autoref{fig:sph} we show the cross section of the most viscous disc ($\alpha=0.2$) at the end of the simulation. As expected, the disc shows a warp in the inner regions. To be more quantitative, we illustrate the tilt and the twist angles as a function of radius for all our simulations in \autoref{fig:tilt}. The angles have been estimated out to $r=30a$, since the disc has significantly spread in the radial direction due to pressure forces (the initial condition truncates the surface density at $r=15a$) and viscous spreading. The tilt dependence on radius does not vary much with viscosity, for a given fixed initial misalignment ($\beta_0$). However, the tilt as a function of radius shows a strong dependence ($\sim$ linear) on $\beta_0$. This indicates that the tilt structure is mostly determined by the non-Keplerian disc profile \citep[see Section 3.3 of][]{2013ApJ...764..106F}. The semi-analytical simulations by \citet{2014MNRAS.442.3700F} also confirm that this should happen in the region of the parameter space explored in this paper. For the highest viscosities, the disc is slowly getting aligned with the binary plane, as expected from viscous damping \citep{2000MNRAS.317..773B,2013MNRAS.431.2655K}. On the contrary, the twist angle does not depend on the the initial misalignment, $\beta_0$. This is also shown in \autoref{fig:tilt}{\it b}, where we plot the difference between the twist angle in the outer and in the inner regions of the simulated disc ($\Delta\gamma$) as a function of viscosity. The angle $\Delta\gamma$ is roughly linear with $\alpha$, as analytically predicted by \citet{2013ApJ...764..106F} and confirmed by \citet{2014MNRAS.442.3700F} (in the bending wave regime).

\subsection{Observability of warp fine structures}
\label{sec:prime_angles}
While the tilt and the twist angles contain important informations on the formation and evolution of warped discs they are observationally extremely 
challenging to determine with current instrumentation. 
The most important reason for that is that $\beta$ and $\gamma$ describe the orientation of the specific angular momentum of the disc at each
spherical radius with respect to that of the central binary. Therefore, to observationally constrain $\beta$ and $\gamma$ one has to know the
orientation of the binary orbit and/or their specific angular momentum in an absolute sense. This is extremely challenging
to measure as either the binary cannot be spatially resolved, or there might be not enough astrometric measurements to precisely determine the
orientation of the orbit. 

Given the lack of information on the binary angular momentum vector, spatially resolved observations of circumbinary discs can only provide us with information in a reference frame that is aligned with the disc itself at a certain reference radius. 
To derive observable quantities in this reference frame we assume a coordinate system whose positive z-axis is aligned with the specific angular 
momentum of the outer edge of the disc. We will denote the quantities in this coordinate system with $^\prime$. Thus $\beta^\prime$ and 
$\gamma^\prime$ will now describe the warp structure with respect to the disc itself. 
$\beta^\prime$ describes the relative inclination of a given annulus in the disc with respect to that of the outer edge of the disc, while
$\gamma^\prime$ defines the direction of the line of nodes of a given annulus with respect to an arbitrary direction in the 
$x^\prime-y^\prime$ plane. Similar to the definition of $\beta$ and $\gamma$, $\beta^\prime$ and $\gamma^\prime$ can be calculated from the specific 
angular momentum at each radius. 
The transformation between the frame of the binary and the frame of the outer edge of the disc is given by 
\begin{equation}
{\bf l^\prime}(x^\prime, y^\prime, z^\prime) = {\bf R}_y(\beta_{\rm out})\cdot {\bf R}_z(\gamma_{\rm out})\cdot {\bf l}(x,y,z) 
\end{equation}
where $\cdot$ represents matrix multiplication, $\beta_{\rm out}$ and $\gamma_{\rm out}$ are the twist and the tilt angles taken at the outer radius of the disc, while ${\bf R}_y(\beta_{\rm out})$ and  ${\bf R}_z(\gamma_{\rm out})$ are rotation matrices around the y- and z-axis, respectively 
\begin{equation}
{\bf R}_y(\beta_{\rm out})= 
\begin{pmatrix}
\cos{\beta_{\rm out}} & 0 & -\sin{\beta_{\rm out}} \\
0 & 1 & 0 \\
\sin{\beta_{\rm out}} & 0 & \cos{\beta_{\rm out}} \\
\end{pmatrix}
\end{equation}

\begin{equation}
{\bf R}_z(\gamma_{\rm out}) = 
\begin{pmatrix}
\cos{\gamma_{\rm out}} & \sin{\gamma_{\rm out}} & 0 \\
-\sin{\gamma_{\rm out}} &  \cos{\gamma_{\rm out}} &0 \\
0 & 0 & 1\\
\end{pmatrix}.
\end{equation}

Thus the normalised specific angular momentum components in the coordinate system aligned with the plane of the outer edge of the disc will be

\begin{eqnarray}
\nonumber l^\prime_x &=& \cos{\beta_{\rm out}}\cos{\gamma_{\rm out}}\sin{\beta(r)}\cos{\gamma(r)} + \\
& & \cos{\beta_{\rm out}}\sin{\gamma_{\rm out}}\sin{\beta(r)}\sin{\gamma(r)} - \sin{\beta_{\rm out}}\cos{\beta(r)}\\
l^\prime_y &=& \cos{\gamma_{\rm out}}\sin{\beta}\sin{\gamma(r)} - \sin{\gamma_{\rm out}}\sin{\beta}\cos{\gamma(r)} \\
\nonumber l^\prime_z &=& \sin{\beta_{\rm out}}\cos{\gamma_{\rm out}}\sin{\beta(r)}\cos{\gamma(r)} + \\
& & \sin{\beta_{\rm out}}\sin{\gamma_{\rm out}}\sin{\beta(r)}\sin{\gamma(r)} - \cos{\beta_{\rm out}}\cos{\beta(r)}.
\end{eqnarray}

The twist and the tilt in this frame of reference can now be calculated from 

\begin{eqnarray}
\cos{\beta^\prime} &=& l^\prime_z\\
\tan{\gamma^\prime} &=& \frac{l^\prime_y}{l^\prime_x}.
\end{eqnarray}

We present $\beta^\prime$ and $\gamma^\prime$ as a function of radius in \autoref{fig:tilt}{\it c,d}. The total tilt of the disc, i.e. the difference in 
the tilt angle at the outer and the inner radii, is very similar both in the frame of the binary ($\Delta\beta$) and in the frame of the disc ($\Delta\beta^\prime$). 
Similar to $\beta$ the $\beta^\prime$ curves also form two groups separated by the initial misalignment angle between the central binary. This suggests that 
in our simulations the initial misalignment angle is the most important parameter determining the tilt of the disc, and this is independent of the choice of the 
frame of reference. Interestingly, $\Delta\gamma^\prime$ is significantly smaller than $\Delta\gamma$.  The total twist in the frame of the disc is not larger 
than 10\degr  in any of the models.

\section{Radiative transfer}
\label{sec:rt}

To study the detectability of warps in protoplanetary discs we used the 3D radiative transfer code \textsc{radmc-3d}\footnote{http://www.ita.uni-heidelberg.de/~dullemond/software/radmc-3d/}. The radiative transfer simulations contained two steps. First the temperature of the dust in the disc
was calculated in a thermal Monte Carlo simulations, then we calculated images in near-infrared scattered light as well as in continuum and gas lines
at sub-millimetre wavelengths.  

We used a 3D spherical mesh to discretise our model with N$_r$=200, N$_\theta$=180, N$_\phi$=180 grid cells in the radial, poloidal and azimuthal
directions, respectively. The inner and outer radius of the mesh was  chosen to be 10\,au and 150\,au and the grid cells were distributed logarithmically
in the radial directions. In the poloidal directions 10,80,80,10 grid cells were distributed in an equidistant grid in the [0, $\pi$/2-1.1], [$\pi$/2-1.1, $\pi$/2], [$\pi$/2, $\pi$/2+1.1], [$\pi$+1.1, $\pi$] intervals, respectively. For the azimuthal grid we used an equidistant grid in the [0, 2$\pi$] interval. 

Since SPH simulations in general do not have high enough resolution high above the mid-plane, 
where the density is significantly lower than in the disc mid-plane (thus the smoothing length is very large), we did not use the density of the SPH 
simulations directly. Instead we used the tilt ($\beta$) and twist ($\gamma$) angles as well as the surface density as a function of radius ($\Sigma(r)$) 
to generate the density structure of the disc in the following way. We described the density structure as 
\begin{equation}
\rho(x^{\prime\prime}, y^{\prime\prime}, z^{\prime\prime}) = \frac{\Sigma(x^{\prime\prime},y^{\prime\prime})}{H_{\rm p}(x^{\prime\prime}, y^{\prime\prime})\sqrt{2\pi}}\exp\left[-\frac{{z^{\prime\prime}}^2}{2H_{\rm p}(x^{\prime\prime},y^{\prime\prime})^2}\right]
\label{eq:disc_dens}
\end{equation}
where $H_{\rm p}$ is the pressure scale height and $\Sigma$ is the surface mass density. $x^\prime$, $y^\prime$ and $z^\prime$ are the cartesian 
coordinates in a frame of reference in which the $x^\prime y^\prime$-plane is aligned with the plane of the outer edge of the disc. $x^{\prime\prime}, 
y^{\prime\prime}, z^{\prime\prime}$ are the cartesian coordinates in the frame in which the $x^{\prime\prime} y^{\prime\prime}$-plane is aligned with the 
plane of an annulus in the disc at a given radius. The transformation between the two frames is given by 

\begin{equation}
{\begin{pmatrix}
x^{\prime\prime}\\
y^{\prime\prime}\\
z^{\prime\prime}\\
\end{pmatrix}}
=
\begin{pmatrix}
\cos{\beta^\prime} & 0 & \sin{\beta^\prime}\\
0 & 1 & 0\\
-\sin{\beta^\prime} & 0 &  \cos{\beta^\prime}\\
\end{pmatrix}
\begin{pmatrix}
\cos{\gamma^\prime} & -\sin{\gamma^\prime} & 0 \\
\sin{\gamma^\prime} & \cos{\gamma^\prime} & 0 \\
0 & 0 & 1\\
\end{pmatrix}
\begin{pmatrix}
x^\prime\\
y^\prime\\
z^\prime\\
\end{pmatrix}
\end{equation}
Here, $\gamma^\prime$ and $\beta^\prime$ are the radial dependent twist and tilt angles, respectively, derived from the SPH simulations. 
The connection between the cartesian coordinates and the spherical coordinates of the spatial grid is given by the usual transformation
\begin{eqnarray}
\nonumber x^\prime &=& r\sin{\theta}\cos{\phi}\\
\nonumber y^\prime &=& r\sin{\theta}\sin{\phi}\\
z^\prime &=& r\cos{\theta}.
\end{eqnarray}

We assumed that the pressure scale height to be a power-law as a function of radius such that
$H_{\rm p}(r) = 0.1(r/r_{\rm in})^\zeta$ with $r_{\rm in}$ being the inner radius of the disc, which we assumed to be at 10\,au, and
$\zeta=0.25$ is the flaring index, in agreement with the sound speed radial profile used in the SPH simulations. The disc outer radius $r_{\rm out}$ is thus at 150\,au.
We assumed that the stars have the parameters of Herbig stars, i.e. R$_\star$=2.5\,R$_\odot$, M$_\star$=2\,M$_\odot$, T$_\star$=9500\,K, 
and a distance of the source from the observer of 100\,pc. Dust particles in the model had a grain size distribution between 0.1\,$\mu$m and 1\,mm with a power exponent of -3.5.
The dust opacity was calculated from the optical constants of astronomical silicate \citep{weingartner_2001} using Mie-theory. 
The disc mass was assumed to be 0.01\,$M_{\odot}$ with a dust-to-gas ratio of 0.01. The stellar parameters as well as the most important parameters of our disc model are summarized in \autoref{tab:rt_summary}. For the CO abundance, with respect to molecular hydrogen, 
we assumed $10^{-4}$ and a $^{17}$O/$^{16}$O isotopic ratio of 2160. In the whole paper, with CO we mean the $^{12}$CO isotopologue. CO and its isotopologues are known to freeze out to dust grains if the dust temperature drops below $\sim19$\,K decreasing the gas phase abundance of these molecules by several orders of magnitude. However, in our models the temperature is above 19\,K everywhere in the disc due to the high luminosity of the stars, thus freeze out of CO and its isotopologues does not have any effect on our models. 
To simulate photodissociation by the UV radiation of the stars we removed all CO from the upper layers of the disc  where the radial optical depth as seen from the stars at 0.5\,$\mu$m is lower than unity.

\begin{table}
\caption{Summary of the stellar and disc parameters used in the radiative transfer calculations.}
\centering
\begin{tabular}{ll}
\hline
M$_\star$ & 2\,M$_\odot$\\
R$_\star$ & 2.5\,R$_\odot$\\
T$_{\rm eff}$ & 9500\,K\\
M$_{\rm disc}$ (dust+gas) & 0.01\,$M_\odot$\\
M$_{\rm dust}$/M$_{\rm gas}$ & 0.01\\
R$_{\rm in} $& 10\,au\\
R$_{\rm out} $& 150\,au\\
H$_p(R_{\rm in})$& 0.1\\
$\zeta$ & 0.25\\
\hline
\hline
\label{tab:rt_summary}
\end{tabular}
\end{table}

We calculate continuum images in the H-band at 1.65\,$\mu$m and 880\,$\mu$m (ALMA Band 7) as well as channel maps in the J=3-2 line of CO 
(345.795989\,GHz). To study the effect of optical depth on the line profile we also calculated channel maps in the J=3-2 transition of $^{13}$CO (330.5879652218\,GHz), C$^{18}$O (329.3305525\,GHz) and C$^{17}$O (337.0611298\,GHz). For the channel maps we assumed a velocity resolution of 0.2\,km/s. To study the effect of inclination we calculate images at 0\degr, 
45\degr and 90\degr inclination angles. Since the inclination angle of the disc changes as a function of radius due to the warping of the disc, 
unless specified otherwise, inclination means the inclination angle at the outer edge of the disc. Synthetic observations were created by convolving the 
near-infrared images with a 0.04\arcsec Gaussian PSF, which is the resolution of SPHERE on VLT, the current state-of-the-art near-infrared camera. 
In the sub-millimetre synthetic observations with ALMA were generated using the Common Astronomy Software Applications (CASA) v4.2.2 using the 
{\tt simobserve} task to generate synthetic visibilities and the {\tt clean} task for imaging. The full 12m Array was used for the simulated observations in 
two different configurations resulting in a spatial resolution of approximately 0.09\arcsec. For continuum simulations we assume the full 7.5\,GHz 
continuum bandwidth of the ALMA correlator and calculate the visibilities for a total integration time of 30 minutes. For the gas line simulations we 
assumed an integration time of 4\,h. 

\begin{figure}
\center
\includegraphics[width=\columnwidth]{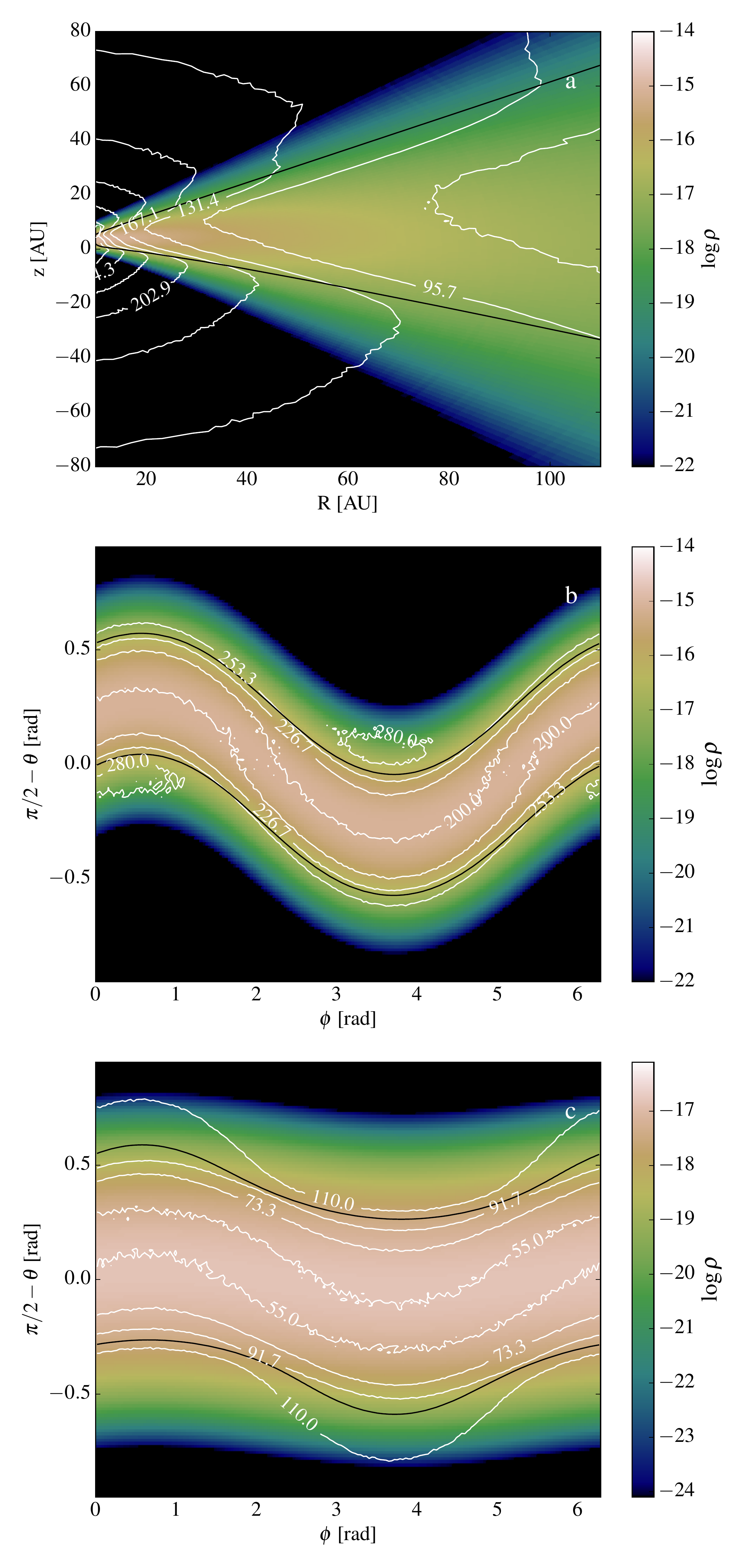}
\caption{Density and temperature structure of a warped circumbinary disc in a vertical slice in the $rz$-plane at $\phi=0^\circ$ ({\it a}) 
and in the $\phi-\theta$ surface at $r=12$\,au ({\it b}) and at $r=100$\,au ({\it c}). The colour-scale image shows the logarithm of the dust density 
in the disc while the white contours show the dust temperature. The black contour shows the location of the radial optical depth of unity at
0.5\,$\mu$m.  
}
\label{fig:discstruct_crossec}
\end{figure}

We did not add any instrumental or atmospheric noise to our synthetic observations. The noise seen in the 
images are numerical noise originating in hydrodynamic and/or radiative transfer calculations or due to the incomplete sampling of the 
Fourier-plane in the synthetic ALMA observations. The reason for the lack of any additional observational
uncertainty in our synthetic observations is that our intention was to show the amount of asymmetry intrinsic to a warped disc without any corruption
by an instrument or the Earth's atmosphere. The noise level in any observation depends on the parameters of the observation itself (e.g. integration
time, instrumental characteristics, weather conditions, etc). If we assumed a given set of such observational parameters the resulting noise level
may hide some of the effects related to the presence of a warp, such as surface brightness asymmetry, which may not show up in our predictions
but would be visible in deeper observations.

While observables are calculated for all hydrodynamic models in the following we show the results of the simulation
with $\alpha=0.2$. To identify the signatures of the warp in the synthetic observations, we also calculated a reference
model without a warp. To ensure a meaningful comparison in the reference model we took the surface density from our fiducial model with $\alpha=0.2$, but we set $\beta(r)$ and $\gamma(r)$ to zero.

\begin{figure*}
\center
\includegraphics[width=\textwidth]{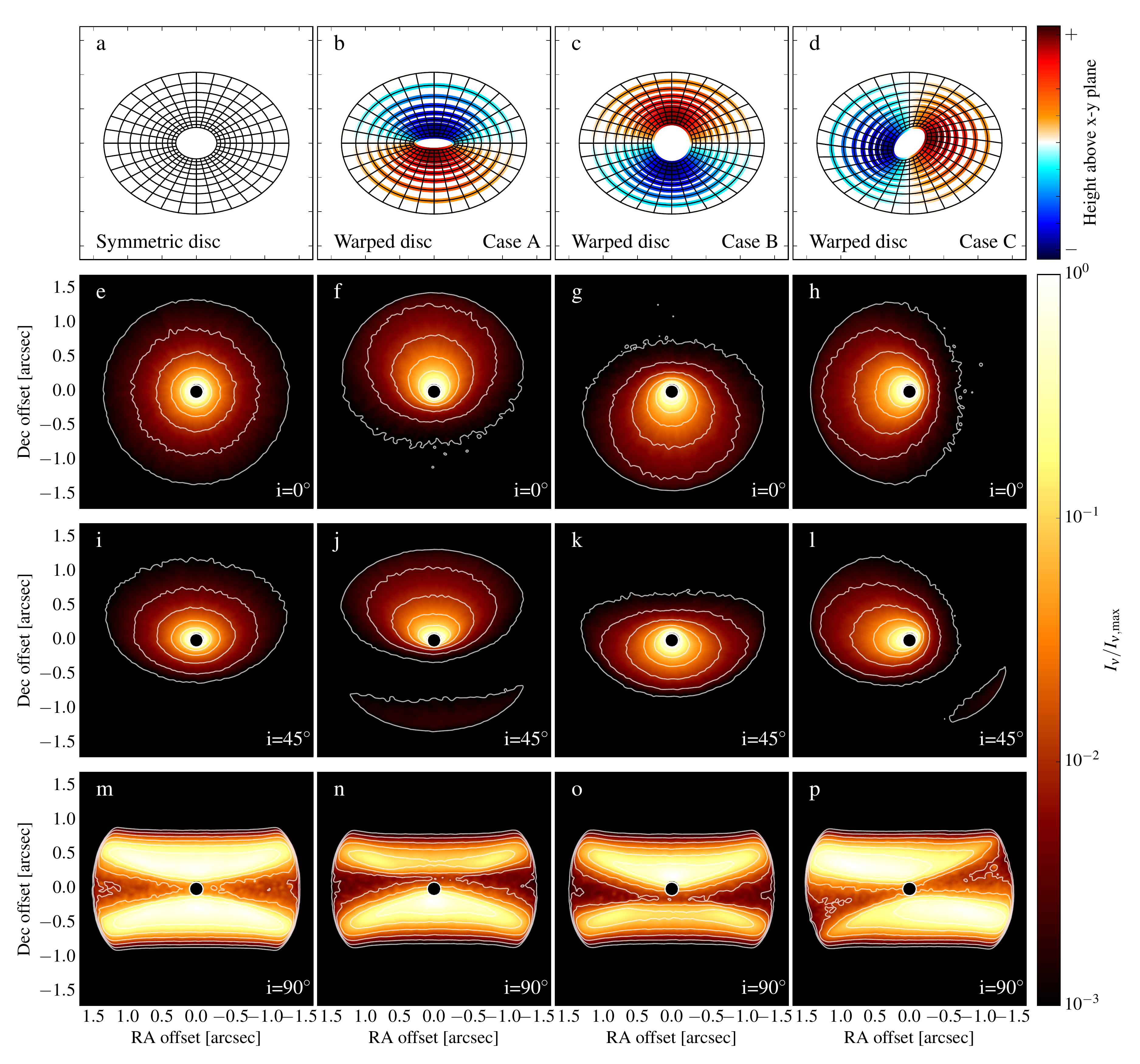}
\caption{Morphology of warped discs in near-infrared scattered light observations. Panel {\it a} shows a cartoon of a symmetric disc while 
Panel {\it b--d} shows a cartoon depicting the structure and orientation 
of the warp shown in the synthetic images in each column. The colourscale shows the height above the mid-plane of the disc with blue showing structures below the mid-plane while red shows regions 
above the mid-plane. The second, third and fourth row from the top show the scattered light images at $i=0^\circ$, $i=45^\circ$ and $i=90^\circ$ inclination
angles, respectively. The second, third and fourth column from the left shows the synthetic H-band scattered light images of our fiducial warped disc model, 
while the leftmost column shows the images predicted images of an unperturbed, symmetric disc. Warped discs show an azimuthally asymmetric surface
brightness distribution at all inclination angles, even if seen perfectly face on. The surface brightness asymmetry is due to the shadowing of the outer disc
 by the inner regions. The position angle of the asymmetry is set by the orientation/position angle of the warp. 
}
\label{fig:image_nirscat}
\end{figure*}

\section{Results}
\label{sec:results}

\subsection{Disc structure}
\label{sec:disc_structure}
We present the disc density and temperature structure of our fiducial model in \autoref{fig:discstruct_crossec}{\it a--c}. 
The temperature structure of a warped disc is significantly altered compared to an unperturbed disc. 
In a symmetric unperturbed passive irradiated disc the vertical density and temperature profiles are inversely correlated such that the temperature 
increases from the disc mid-plane, where the density is the highest, towards the disc atmosphere, where the density is the lowest. In a warped disc this 
is not necessarily the case. While close to the inner edge of the disc this inverse correlation still holds (\autoref{fig:discstruct_crossec}{\it b}), it
breaks further out in the disc. It can be seen in  \autoref{fig:discstruct_crossec}{\it c} that at 100\,au distance the vertical temperature profile is not 
coupled to the local density, but shows correlation (i.e. similar azimuthal modulation) with the vertical density structure in the inner disc at 12\,au. 
This means that at 100\,au radius, the lowest temperature occurs at different heights and densities at different azimuth angles. 
Since in a passive disc the dust temperature is determined by the absorbed stellar radiation, the correlation between the density at the inner edge
of the disc and the temperature at larger distances from the star suggests that the primary cause of the 
temperature perturbation is the self-shadowing of the disc by its inner edge where it is curved out of the plane of the outer disc. 

The requirement for the inner disc to cast a shadow at a given radius $r$ is 
$\beta^\prime(r_{\rm in})\geq H_{\rm s}(r)/r-H_{\rm s}(r_{\rm in})/r_{\rm in}$, where $H_{\rm s}(r)$  the height of the surface layer above the mid-plane. 
In the unperturbed, reference model   $H_{\rm s}(r)/r=0.23$ and $H_{\rm s}(r)/r=0.38$ at the inner and outer radius of the disc, respectively. 
Thus we would expect to see the effect of non-axisymmetric self-shadowing if $\beta^\prime(r_{\rm in})\geq 8.6^\circ$, which is the case in all but
two models (see \autoref{fig:tilt}). Even though this limit may change somewhat depending on the optical properties of the dust grains in the disc 
as well as on the flaring index, it is reasonable to assume that the critical $\beta^\prime(r_{\rm in})$ above which the non-axisymmetric self-shadowing
occurs is on the order of $H_{\rm p}/r$.

Due to the strong three dimensional perturbations one expects two kinds of observational signatures from a warped circumbinary disc: surface
brightness asymmetries caused by the self-shadowing of the disc and kinematical asymmetries due to the globally non-axisymmetric velocity field. 
In the following we will investigate both continuum and gas line observations and look for this kind of observational signature which can
help to identify a warp in an observed disc.

Since a warped disc is a genuine three dimensional structure the signatures depend not only on the inclination of the disc, but also on the 
azimuth angle. Thus we present synthetic observables of warped discs at three different orientation of the disc depending on the azimuth angle, 
measured in the plane of the outer edge of the disc. The three cases can be described as follows:

\begin{description}
\item{\bf Case A:}  The disc bends along the line of sight such that the angle between the direction of the observer and the specific angular momentum of the disc increases with decreasing radius (see Fig.~5\,{\it b}). This means that if the disc is viewed at a non-zero inclination angle the
effective inclination of the disc increases with decreasing radius. 
\item{\bf Case B:} The disc bends along the line of sight such that the angle between the direction of the observer and the specific angular momentum of the disc decreases with decreasing radius (see Fig.~5\,{\it c}). If the disc is viewed at a non-zero inclination angle the effective
inclination of the disc decreases with decreasing radius. Case B can be obtained from the position of Case A by of rotating the disc by 180\degr 
around the rotation axis of the outer edge of the disc. 
\item{\bf Case C:} The disc bends perpendicular to the line of sight  (see Fig.~5\,{\it d}). The warping of the disc is such that the projected specific angular momentum of the disc onto the plane of the sky is rotated counter-clockwise with decreasing radius. Case C can be obtained from Case A
by rotating the disc counter-clockwise by 90\degr or from Case B by rotating the disc clockwise around the rotation axis of the outer edge of the disc.
\end{description}

\subsection{Continuum images}

Due to the lack of kinematical information, in continuum images the observable signatures of disc warps is limited to surface brightness asymmetries.
Variation in the observed intensity can be caused by the variation of the density, of the temperature or a combination of both. The wavelength of
observation and the local dust temperature will determine which physical parameter causes the surface brightness asymmetry.

\subsubsection{Near-infrared images}
\label{sec:NIR_continuum_images}

\begin{figure*}
\center
\includegraphics[width=\textwidth]{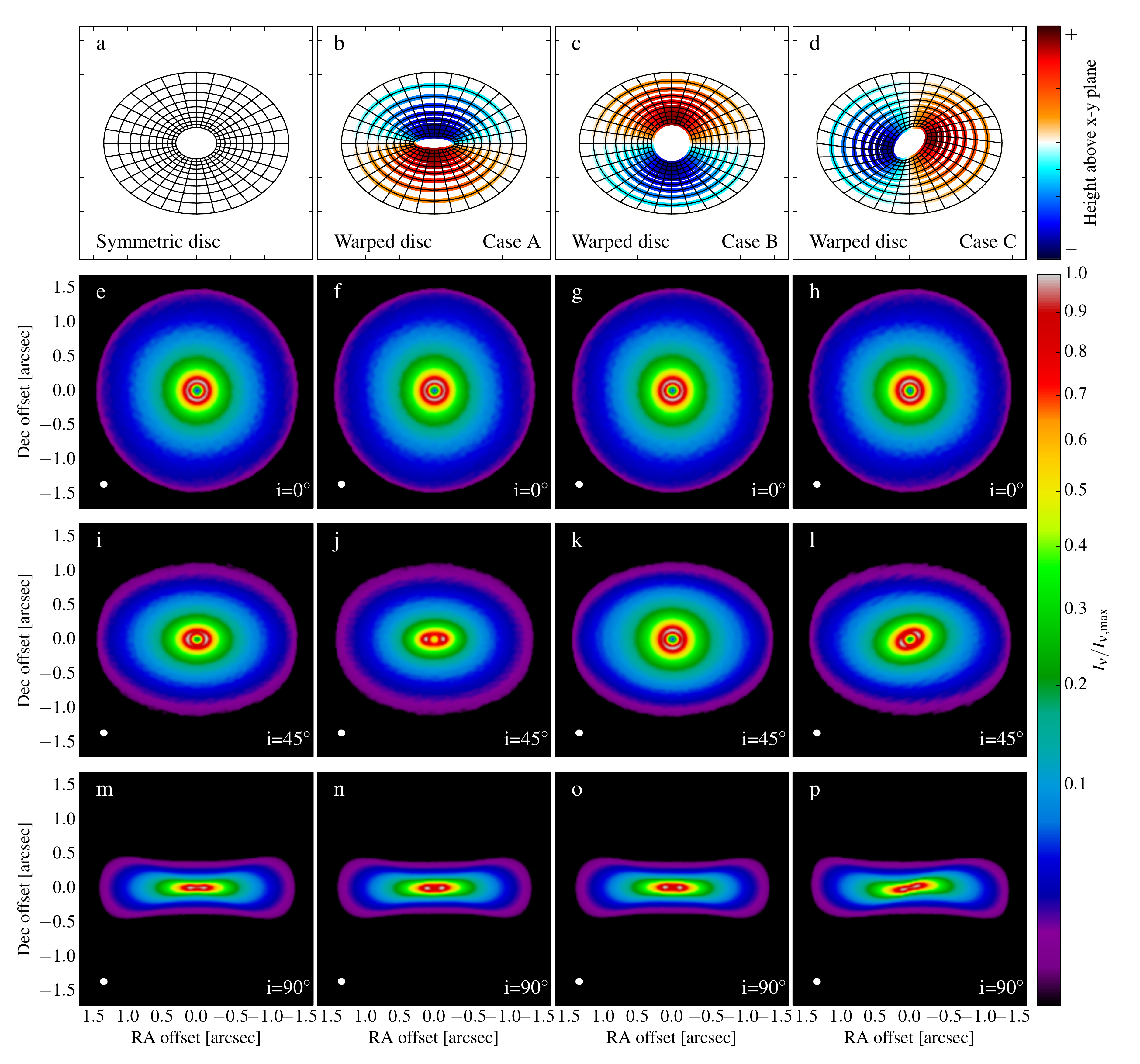}
\caption{Same as \autoref{fig:image_nirscat}, but the colorscale images show synthetic ALMA observations at 340\,GHz. The synthesised beam is shown in white in the bottom left corner of each panel. No obvious azimuthal asymmetry is visible
in the sub-millimetre continuum images of warped discs.  The isophotes of an inclined disc are concentric ellipses, whose position angle and aspect
ratio may change as a function of radius. The strength of this effect increases with increasing inclination angle. }
\label{fig:image_submm_cont}
\end{figure*}

Current state-of-the-art NIR cameras (e.g. SPHERE/VLT, GPI/Gemini, HiCIAO/SUBARU) can probe regions in the disc down to several au distances
from the central star in nearby star-forming regions. At these radii the temperature in the disc is too low for significant thermal emission at NIR 
wavelengths. Thus NIR observations, dominated by scattered stellar radiation, probe mainly density variations in the upper layers of the disc. 

For face-on discs, the NIR image of an unperturbed disc shows a circular symmetric surface brightness distribution (see \autoref{fig:image_nirscat} e). A 
warped disc on the other hand shows a lopsided brightness distribution such that the disc can be divided azimuthally into a bright illuminated and a dark shadowed 
region each having approximately the same azimuthal extent of about 180\degr.
In the fiducial model with $\alpha=0.2$ the peak-to-peak variation of the surface brightness at r=0.5$^{\prime\prime}$ is a factor of 12 while it is a factor of 16 at r=1$^{\prime\prime}$ for face-on orientation. It is not only the absolute surface brightness that is different in a warped disc compared to a symmetric disc, but
also the radial power exponent of the surface brightness distribution. In our symmetric control model the radial surface brightness falls off as $R^{-2.5}$, where
$R$ is the distance from the central star in the plane of the sky at face-on inclination. In our fiducial warp model the radial surface brightness show strong
azimuthal variation. Fitting a power law to the radial surface brightness profile at the position angle of the highest and lowest surface brightness results in a 
$R^{-3.0}$ and $R^{-2.2}$ profile, respectively. 

We also note that the position angle of the lowest/highest surface brightness depends on the azimuth angle of the warp. The strong asymmetry in the 
surface brightness as well 
as its variation with the azimuth angle of the warp can be understood in terms of the self-shadowing of the outer regions of the disc by the inner disc due to 
the warping. In the bright side the curvature of the disc is such that it allows stronger irradiation of the disc, i.e. it curves upwards if we are looking at the
disc from the top. In the dark side, on the other hand, the situation is the opposite, the outer disc curves downwards into the shadow of the inner 
regions.  
This explanation is further supported by the fact that the amplitude of the surface brightness asymmetry depends on the strength of the warp, more precisely on $\Delta\beta^\prime$. The peak-to-peak variation of the surface brightness in the azimuthal direction is only about a factor of 2.5 in case
of the weakest warp ($\alpha=0.02$ and an initial misalignment of 15\degr,  see \autoref{fig:app_image_nirscat}). For the weakest warp in 
our hydrodynamic simulations, $\beta^\prime(r_{\rm in})=5.9^\circ$, which is very close to $H_{\rm s}(r_{\rm in})/r_{\rm in}=0.1$. This is a lower value
than the criterion we derived in Section~\ref{sec:disc_structure} for $\beta^\prime(r_{\rm in})$ above which the effect of non-axisymmetric shelf-shadowing 
is visible. The reason for this is that the azimuthal asymmetry on the dark side of the disc depends on the self-shadowing, however, on the opposite site
the curvature of the disc is such that the incident angle of the stellar radiation is higher than in the unperturbed disc, which makes that side of the disc brighter, 
making the overall contrast higher than we estimated before from the self-shadowing alone.

For intermediate inclinations (see  \autoref{fig:image_nirscat} i-l) the surface brightness shows similar characteristics as for face-on discs, meaning
that one side of the disc is brighter than the other. Interestingly in a warped disc the position angle of the dark side is determined by the azimuth angle
of the warp. This is in contrast to the unperturbed disc, where the asymmetry is always between the near and far side. In an unperturbed disc the far
side of the disc looks spatially more extended due to the projection, but this can change due to anisotropic scattering. For strongly forward peaking
scattering phase functions the near-side of the disc might be brighter. This suggests a caution when determining the inclination and the position angle 
of a warped disc from NIR images. Another interesting feature is an arc-like structure visible beneath the ellipse of the disc as seen in 
\autoref{fig:image_nirscat}j and \autoref{fig:image_nirscat}l. This arc is caused by photons scattered from the lower side of the disc. 
In \autoref{fig:image_nirscat}j the disc on the near side curves downwards, so that the outer regions in the upper near side of the disc are completely in the shadow of the inner parts. In the lower side of the disc, however, the disc curves outwards of the shadow of the inner regions making it more
exposed to direct stellar radiation. 

NIR images of edge-on discs show two characteristic bright horizontal lanes, caused by scattering of photons in the
optically thin disc atmosphere (see \autoref{fig:image_nirscat}m). These two bright arcs are separated by a dark lane in the middle, caused by heavy 
extinction in the disc mid-plane. Since the structure of the bright and dark lanes, i.e. the signature of the disc atmosphere and the mid-plane, 
respectively, is determined by the vertical structure of the disc, warped discs are expected to show an asymmetry even at such high inclination. Indeed, 
as can be seen in \autoref{fig:image_nirscat}n-p, the asymmetry is such that the two bright horizontal lanes will not be symmetric to the dark lane, 
but either of them will be broader and brighter than the other. If the warp is perpendicular to the line of sight (see \autoref{fig:image_nirscat}p) the 
dark lane of the disc midplane will show a tilt with respect to the bright lanes of the disc atmosphere. In other words the bright lanes will show a left-right 
asymmetry. 

It is widely known that asymmetries in the brightness distribution of circumbinary and circumstellar discs can be caused by anisotropic scattering
if large grains (comparable or large than the wavelength of observations) are present in the disc atmosphere. Disentangling between
the two mechanisms, anisotropic scattering vs. warp, is in most cases not difficult. Anisotropic scattering causes a surface brightness asymmetry, 
that is always along the minor axis of the projected disc ellipse and the disc seems to be symmetric along the major axis. The strength of the 
asymmetry is inclination angle dependent, it increases with increasing inclination. In contrast, azimuth angle of the surface brightness asymmetry 
caused by a warp is determined by the azimuth angle of the warp which in general is unlikely to be perfectly aligned with the minor axis of the disc.
Moreover, the strength of the surface brightness asymmetry of a warp is independent of the inclination as it is determined by the shadowing, i.e. by 
the maximum tilt at the inner edge of the disc. Therefore disentangling between the effect of anisotropic scattering and a warp becomes easier
for lower inclination angles.

\begin{figure*}
\center
\includegraphics[width=\textwidth]{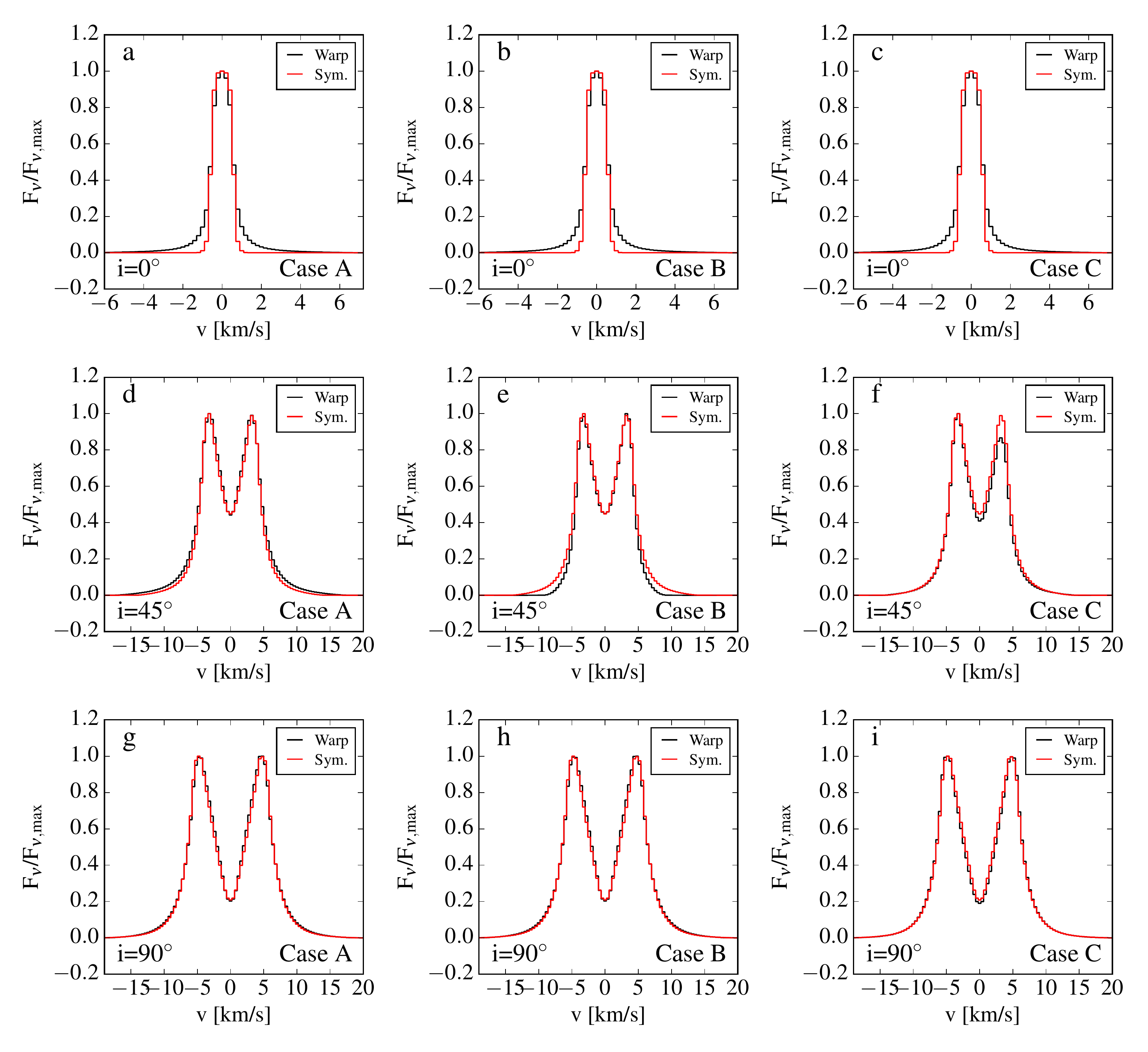}
\caption{Spatially integrated line profile of the CO J=3-2 line of our fiducial model. The black line shows the spectrum of a warped disc while the red line
shows the line profile of a symmetric unperturbed disc for comparison. The inclination and position angle of the warp is shown in the bottom left and 
right corners, respectively. For low inclination angles (Panel {\it a,b,c}) the line profile of warped disc shows extended wings, indicating 
significantly higher projected velocities close to the inner edge of the disc than in an unperturbed disc. 
For face-on discs the presence and strength of the extended wing component is 
independent of the orientation of the warp. At intermediate inclination angles (Panel {\it d,e,f}) the presence of an excess or lack of emission on 
the line wings depends on the position of the warp, i.e. whether the effective inclination angle increases or decreases inwards in the disc. 
If the disc orientation is such as in Case C, i.e. if the disc bends close to perpendicular to the line of sight the surface brightness asymmetry in the 
disc causes an asymmetry in the strength of the Keplerian peaks (see Panel {\it f}). The strength of the deviation of the line profile from that of an 
unperturbed disc decreases with increasing inclination. For edge-on discs (Panel {\it g,h,i}) the line profile of a warped disc is practically indistinguishable 
from that of an unperturbed disc. 
}
\label{fig:specshape}
\end{figure*}

\subsubsection{Sub-millimetre continuum images}
\label{sec:submm_continuum_images}
Protoplanetary discs, around binaries as well as around single stars, are mostly optically thin in the sub-millimetre continuum, except for the very inner regions of the disc (see e.g. \citealt{ricci_2012}), therefore the observed intensity is sensitive to the variation of both column density and temperature. 
Interestingly the derived maps from synthetic 340\,GHz continuum observations of our fiducial warp model are largely symmetric and do not show the large scale surface-brightness asymmetry we see in the near-infrared (see \autoref{fig:image_submm_cont}). The surface brightness variation in the azimuthal direction is not larger than about 10--15\,\% for face-on inclination. 
The reason for that is most likely a combination of the low optical depth of the disc at this wavelength and the fact that the structure of the warp is point-symmetric with respect to the centre of the disc.  The point-symmetry of the disc structure means that if the upper side of the disc is warmer than the lower side at any position in the disc, in the diagonally opposite position the lower side will be warmer
than the upper side by the same amount. Due to the optical thinness of the disc, all layers contribute to the observed emission, therefore the total intensity integrated along the line shows very little variation as a function of azimuth angle.

However, the presence of a warp in the disc might be inferred in discs at intermediate inclination angles by looking at the radial
variation of the isophotes. The isophotes in sub-millimetre continuum images of an unperturbed, symmetric disc at nonzero inclination angles 
are concentric ellipses with the same eccentricity and position angle of the semi-major axis (see \autoref{fig:image_submm_cont}{\it i}). In case of a warped disc the isophotes will still be ellipses but their eccentricity and the position angle of the semi-major axis will depend on the distance from the star (see \autoref{fig:image_submm_cont}{\it j-l}). In case the disc curves along the line of sight  (see the middle two columns of \autoref{fig:image_submm_cont}) the effective inclination angle of the disc changes as a function of radius. The position angle of the isophote ellipses will
not change with the distance from the star but their eccentricity increases or decreases depending on the curvature of the inner disc. 
If the azimuth angle of the warp is different from the line of sight not only the eccentricity of the isophotes but also their position angle will change
as a function of distance from the central star. 

The change of the morphology of the isophotes as a function of the distance from the star is caused by projection. Due to the optical thinness of the
disc the sub-millimetre emission is dominated by a vertically thin layer close to the mid-plane of the disc due to the vertical Gaussian density 
structure of the disc. Since in the optically thin limit the observed intensity depends linearly on the surface density, temperature and dust opacity, 
the isophotes will mark the position of points in the disc where these physical parameters have the same, or very similar values in the disc mid-plane. 
In an unperturbed disc (right column in \autoref{fig:image_submm_cont}{\it e,i,m}) these parameters vary in an axisymmetric way, thus the isophotes 
will be concentric circles for zero inclination angle and concentric ellipses for nonzero inclination angles, due to the projection of the disc onto the plane 
of the sky. However a warped disc is not a truly axisymmetric structure and the highest density layer of the disc, i.e. the disc "mid-plane" is not a plane 
anymore. The density of a warped disc depends most importantly on the {\it spherical radius} thus the isophotes will be ellipses when projected onto
the plane of the sky. The eccentricity and position angle of the isophotes will however be determined by the local tilt and twist angle. 

\begin{figure}
\center
\includegraphics[width=\columnwidth]{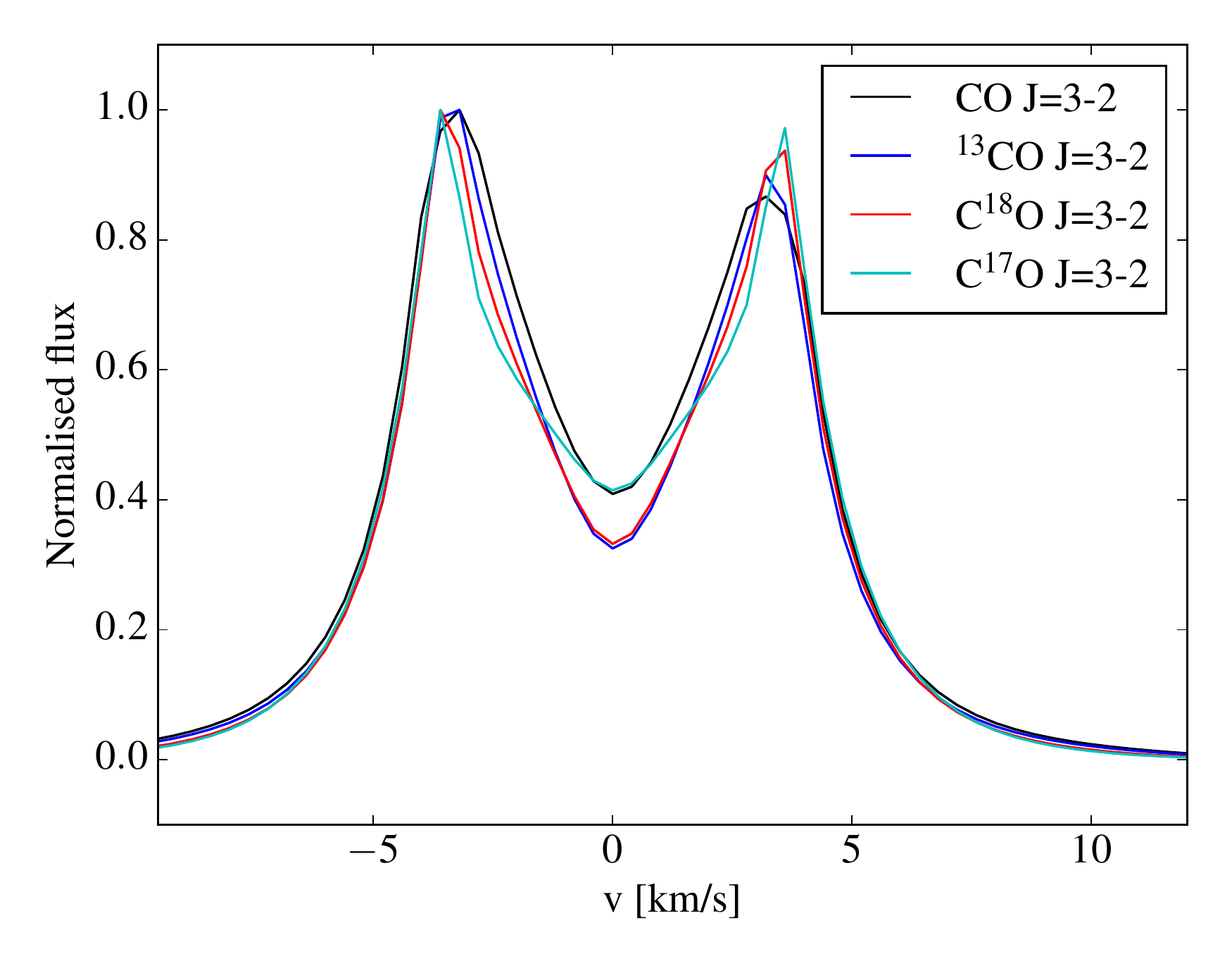}
\caption{Variation of the line profile as a function of optical depth in our fiducial model as seen in the J=3-2 transition of various CO isotopologues. A linear continuum has been subtracted from the spectral lines, which were then normalised to their peak value. The inclination of the outer disc is $i=45^\circ$ and the disc is  bending perpendicular to the line of sight (Case C).  The CO line shows a clear asymmetry in the strength of the Keplerian peaks but the amount of asymmetry in the line profile decreases from CO, through $^{13}$CO and C$^{18}$O to
C$^{17}$O. This behaviour is caused by the increasing azimuthal temperature variation in the disc with increasing height above the midplane. 
The higher the optical depth in the lines, thus the higher they originate above the midplane, the higher the asymmetry of the Keplerian peaks will be. 
} 
\label{fig:lineprof_vs_optical_depth}
\end{figure}

At extreme high inclination angles an asymmetry in the surface brightness distribution is clearly visible if the disc curvature is perpendicular to the line of sight 
(Case C, see \autoref{fig:image_submm_cont}{\it p}). In this case the surface brightness distribution will show an S-shaped curvature along the disc plane. 
This curvature of the surface brightness distribution is 
caused by the projection of the density of the warped disc onto the planet of the sky, thus the curvature of the surface brightness shows the curvature
of the disc plane. If the orientation of the warp is such that the disc curves along the line of sight (see \autoref{fig:image_submm_cont}{\it n, o}) then
the signature of the warp is extremely challenging to see in sub-millimetre continuum observations. At our chosen spatial resolution, high enough to 
resolve the inner hole of the disc, the image of the warped disc cannot be disentangled from that of the unperturbed disc if the curvature of the 
warp is along the line of sight even for the strongest warp in our models. The strength of the signature of the warp in sub-millimetre continuum 
observations decreases with the strength of the warp, i.e. $\Delta\beta^\prime$. The most weakly warped disc in our hydrodynamic simulation can only
be distinguished from an unperturbed disc if the disc bends perpendicular to the line of sight and the inclination is at least $\sim45^\circ$ (see 
\autoref{fig:app_image_submm_cont}). Even in this case the warp signature is very weak thus one would need extremely high S/N to be able to detect it.

\begin{figure*}
\center
\includegraphics[width=\textwidth]{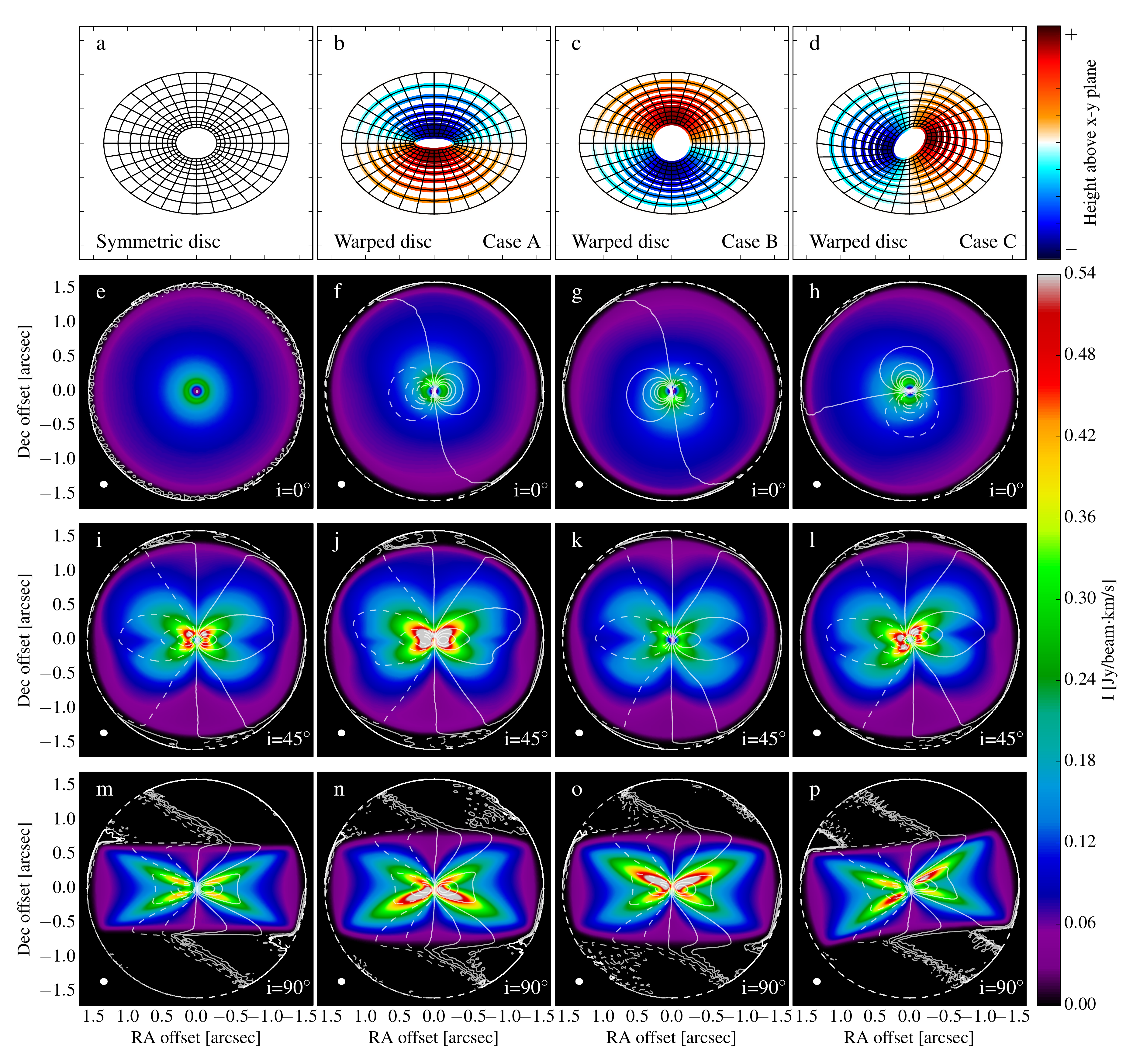}
\caption{Predicted ALMA observations in the CO J=3-2 line. The colourscale images show the integrated intensity (0th moment map) while the
contour lines show the intensity weighted velocity (1st moment) map. Dashed contours denote negative velocity. 
Panel {\it a--d} show the orientation of the warp in the disc in each column. The synthetised beam is shown in white in the bottom left corner of each 
panel. The surface brightness distribution of warped discs shows asymmetries at all inclination angles. For face-on discs (Panel {\it f--h}) the projected
velocity increases towards the inner edge of the disc independently of the orientation of the warp. At intermediate inclinations (Panel {\it j--l}) the 
projected velocity of a warped disc increases or decreases inwards depending on the orientation of the warp. In case the disc does not bend
along the line of sight the velocity field shows a characteristic twist around the centre of the disc (see Panel {\it l}). 
The strength of the kinematical signatures of warp decreases with increasing inclination angle. For edge-on discs the velocity field is practically
indistinguishable from that of an unperturbed disc (Panel {\it n--p}).}
\label{fig:mom1azimut}
\end{figure*}

\subsection{Gas line observations}
\subsubsection{Spatially integrated line profiles}
\label{sec:line_profiles}
In contrast to continuum observations gas lines provide information on both the surface brightness distribution and on the kinematics of the disc.
Kinematical information is especially valuable when it comes to the detection of warps. In \autoref{fig:specshape} we present the spatially
integrated CO J=3-2 spectra of our fiducial warp model derived from the synthetic ALMA observations. The line profiles of warped circumbinary discs 
can show two kinds of deviations from that of an unperturbed, symmetric disc. 

The excess or deficiency of flux in the line-wings compared to a 
profile of a symmetric, Keplerian disc is caused by the kinematical asymmetry alone. If the disc curves along the line-of-sight, the variation of $\beta^
\prime(R)$ causes the effective inclination angle of the disc to change as a function of radius. It can be seen in \autoref{fig:specshape}
that depending on whether the inner disc has a higher (\autoref{fig:specshape}{\it a, d} and {\it g}) or lower inclination angle (\autoref{fig:specshape}
{\it b, e} and {\it h})  than the outer disc, we will see an excess or deficiency of emission at high velocities compared to an unperturbed disc. 
For perfectly face-on orientation, the highest velocity at which we can detect emission ($v_{\rm max}$) depends on $\beta^\prime$ only such that $v_{\rm 
max}=v_{\rm in} \sin{\beta^\prime}$. For non-zero inclinations the expression for $v_{\rm max}$ is more complicated as it depends not only on 
$\beta^\prime$ but also on the inclination angle and on the azimuth angle of the warp at the inner edge of the disc. Nevertheless it is clearly visible
in \autoref{fig:specshape} that the signatures of kinematical deviations from the  line-profile of an unperturbed disc decrease with increasing inclinations.
For edge-on orientation there is barely any difference between the line profile of our symmetric control disc model and the line profiles of a warped
disc. 

The other type of deviation of the line profile of warped discs from that of an unperturbed disc is related to surface brightness asymmetry. Surface 
brightness asymmetries between the left and the right side of the disc translate to an asymmetry of the two Keplerian peaks in the line-profiles of
discs seen at non-zero inclination angle (see \autoref{fig:specshape}{\it f}). The difference in the flux in the two Keplerian peaks in the CO J=3-2
line is on the order of 15--20\,\%. The amount of asymmetry depends also on the inclination (decreases with increasing inclination angle), on the orientation of the warp as well as on the optical depth of the line. The asymmetry is the strongest if the disc curves perpendicular to the line of sight. 
Since the surface density distribution remains symmetric in our models, the asymmetry must be related to the asymmetries in the temperature.  
Indeed, the J=3-2 transition of CO originates high above the mid-plane of the disc, close to the disc atmosphere. 
As we showed in previous Sections the asymmetry in the temperature structure of a warped disc is the strongest in the 
upper layers of the disc and it decreases towards the disc mid-plane. Therefore, the higher above the mid-plane a line originates the stronger the 
asymmetry becomes.  

In \autoref{fig:lineprof_vs_optical_depth} we compare the line profiles of the various CO isotopologues in the J=3-2 transition. As can be seen in
\autoref{fig:lineprof_vs_optical_depth} the amount of asymmetry of the line profile decreases from CO through $^{13}$CO, C$^{18}$O to C$^{17}$O. While the CO line is clearly asymmetric in the strength of the Keplerian peaks, there is barely any asymmetry visible in the C$^{17}$O J=3-2 line.
The reason for that can be found in the region where these lines originate. The highest optical depth across the
CO J=3-2 line is $\sim$4700 indicating that the CO emission originates high above the mid-plane. Since the line is highly optically thick the surface
brightness variation in the CO line reflects the azimuthal variation in the temperature, which we showed earlier to increase with the height above the 
disc mid-plane. While the optical depth of the $^{13}$CO line ($\sim$65) is lower than that of CO it is still highly optically thick and thus its line
profile shows a similar asymmetry to that of the CO line. Even though the highest optical depth in the C$^{18}$O and C$^{17}$O  are higher 
than unity, 4.5 and 2.3, respectively,  they are low enough that the emission is dominated by regions close to the midplane.
Since the variation of the temperature is lower close to the disc midplane than in the layer
where the CO emission originates, the surface brightness -- thus the line-profile -- asymmetry is also higher in the CO line than it is in that of the more optically thin 
isotopologues. We note that in typical protoplanetary discs around T\,Tauri stars the C$^{17}$O J=3-2 is expected to be highly optically thin, due to the freeze out 
of the molecules close to the mid-plane decreasing its abundance. However, in our model the temperature is higher than the freeze out temperature of  CO 
($\sim$19\,K) due to the high stellar luminosity causing the increase of column density, thus optical depth of all CO lines.

The dependence on the asymmetry on the position angle of the warp is also easily explained as well. Asymmetry in the strength of the
Keplerian peaks in the line-profile requires surface brightness difference between the left and the right side of the disc. Therefore this kind of line-profile
asymmetry is the strongest if the warp is perpendicular to the line of sight. In case the disc curves parallel to the line-of sight, the surface brightness
asymmetry due to the self-shadowing of the disc will also be along the line of sight, thus the Keplerian peaks will remain symmetric. 
The strength of the kinematical signature as well as the asymmetry in the line profile depends on the strength of the warp, $\beta^\prime(r_{\rm in})$,
which in turn in this regime is controlled by the initial misalignment angle between the binary and the disc and the viscosity. Indeed in the most weakly
warped model ($\beta^\prime(r_{\rm in})\approx H_{\rm p}/r$) the line profile of the CO J=3-2 line is practically indistinguishable from that of an unperturbed 
disc at most inclination angles. The difference in the strength of the Keplerian peaks for such weakly warped discs is only a few percent (see \autoref{fig:app_specshape}{\it d--i}). The excess emission at high velocities is only at 10\,\% level at face-on inclination and 
it decreases with increasing inclination.

\begin{figure*}
\center
\includegraphics[width=\textwidth]{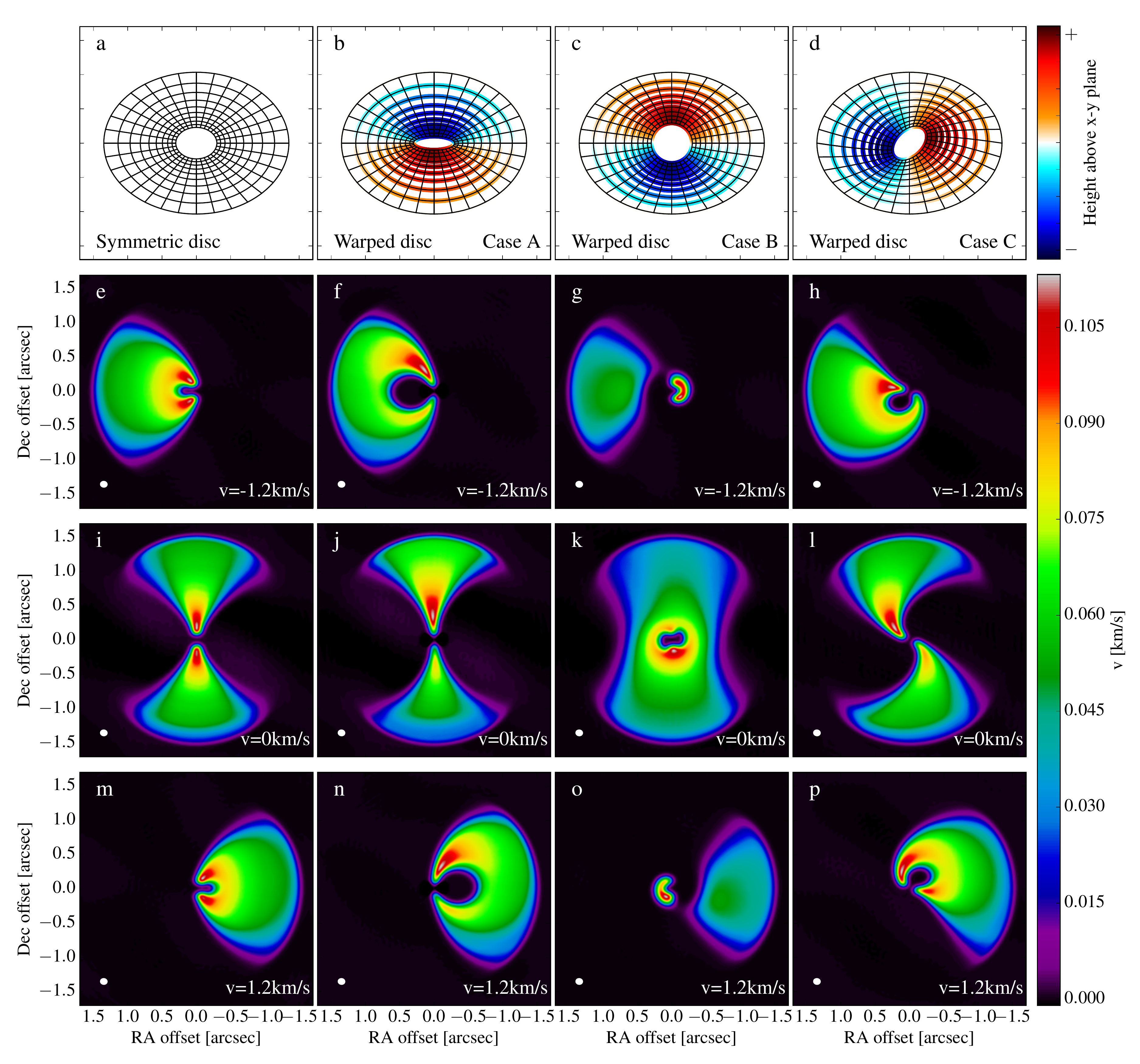}
\caption{Predicted channel maps of warped disc in the CO J=3-2 line in ALMA observations. Panels {\it a--d} shows a cartoon depicting the structure 
and orientation of the warp. The left column shows the channel maps of an unperturbed disc for reference. The channel maps are shown at 
v=-1.2\,km/s (Panel {\it e--h}),  v=0\,km/s (Panel {\it i--l}),  v=+1.2\,km/s (Panel {\it m--p}),  respectively. The inclination of the outer disc is $i=10^\circ$.
 }
\label{fig:channel_maps}
\end{figure*}

We wish to emphasise that it is extremely challenging if not impossible to infer the presence of a warp from spatially integrated line profiles.
The spatially integrated line profile depends on the inner and outer radius of the disc, the inclination angle, and the temperature
profile. The presence of a warp can only be unambiguously determined from spatially integrated line profiles if the inner and outer radius of the disc, 
the inclination angle, and the temperature profile of the disc is known with reasonable accuracy. 
In case of a warped disc, the excess or deficit of emission on the line-wings is caused by the radial dependent effective inclination due to the curvature 
of the disc, that changes the projected velocity as well as the projected emitting area. However, when comparing an observed line profile to a model, 
the assumption of a smaller or larger inner disc radius than in reality can result in an apparent excess or deficit of flux on the line wings in the
observed line compared to the model, mimicing the effect of a warp. A similar effect might be achieved if the temperature profile is steeper or shallower 
in the model than it is in reality and the line is optically thick. In this case the line intensity will change faster or slower with radius than in reality 
affecting the emission on the line wings. 
In a warped disc the asymmetry of the Keplerian peaks is caused by the non-axisymmetric self-shadowing of the disc. 
However, foreground extinction, pointing errors, the presence of another unresolved source at a a slightly different systemic velocity can all cause an 
asymmetry in the strength of the Keplerian peaks. Moreover, there can also be other physical asymmetries in the system, e.g. if the disc becomes eccentric 
due to a companion (see \citealt{regaly_et_al_2010,2014ApJ...785L..31R}), that can cause an
asymmetry between the Keplerian peaks, which are not related to warps.

\subsubsection{Spatially resolved line observations}
\label{sec:moment_maps}
Spatially resolved observations are invaluable in identifying warps in circumbinary discs, especially in case of gas line observations. 
In \autoref{fig:mom1azimut} we present the synthetic CO J=3-2 integrated intensity and intensity weighted velocity maps, i.e. the 
0th and 1st moment maps, respectively. The integrated intensity maps show deviations from that of a symmetric disc at practically all inclination
angles. 

For face-on orientation (see \autoref{fig:mom1azimut}{\it e--h}) the morphology of the azimuthal surface brightness asymmetry is similar to that in NIR scattered light. The surface
brightness on the brighter side of the disc is about 40\,\% higher than on the opposite side of the disc. This asymmetry is caused by the azimuthal 
variation of the temperature in the disc at the height above the midplane where the CO emission originates.  This explanation is also supported by 
the fact that the position angle of the bright and dark sides of the disc is set by the orientation of the warp. 
The intensity weighted velocity map shows no significant radial velocity close to the outer edge
of the disc as expected for a face-on disc. In contrast, close to the inner edge of the disc significant projected radial velocity is detected revealing
the typical projected velocity pattern of a rotating disc at non-zero inclination angle. The lowest and highest velocity detected in our fiducial model 
at the inner edge of the disc is $-5.8$\,/km/s and $+5.8$\,km/s, respectively. This behaviour is caused by the curvature of the disc mid-plane, as
warping of the disc appears as a radial dependent inclination of the disc. Therefore, even though the outer edge of the disc is seen  perfectly
face-on, the inner edge has a non-zero inclination angle due to the warping.

At an inclination angle of $i=45^\circ$, there are two kinds of deviation in the surface brightness distribution of a warped disc from that of an 
unperturbed disc. If the orientation of the warp is such that the disc curvature is along the line of sight (see \autoref{fig:mom1azimut}{\it j,k}) 
The radial surface brightness distribution is steeper or shallower, compared to that of a symmetric disc,  depending on whether the local effective 
inclination angle increases or decreases inwards in the disc. If the inclination angle of the disc increases inwards the projected radial velocity
will increase above that of a symmetric disc or vice versa if the inclination of the disc decreases inwards. In case the disc curvature is perpendicular 
to the line of sight the both the surface brightness distribution and the projected velocity show and "S"-shaped, twisted structure.
This "S"-shaped twist in the velocity map is even more clearly seen at low inclination angles in channel maps at zero velocity (see 
\autoref{fig:channel_maps}). The formation of this
twisted projected velocity map can be easily understood. The zero projected velocity in a symmetric disc is always along the minor axis of the projected disc
ellipse, i.e. along a straight line. However, if the disc is tilted at any given radius perpendicular to the line of sight, the projected velocity along that 
straight line will not be zero anymore. The zero velocity at that particular radius will be shifted to either side of the disc. Thus if the local tilt angle
changes as a function of radius it will cause a radial dependent azimuthal shift in the projected velocity, i.e. it will cause the twisting of the projected
velocity map around the centre of the disc. 

Asymmetry in the surface brightness distribution in our fiducial model is observable in most cases even if the disc is seen edge-on. In Case A/B orientation 
of the warp the asymmetry is between the upper and the lower side of the disc. In this case either the lower or the upper side of the disc gets 
more irradiation than the other, depending on which way the inner disc bends, due to to self-shadowing by the inner disc, which translates to 
temperature asymmetry between the two sides of the disc that is reflected in the optically thick CO emission maps (see Section~\ref{sec:disc_structure} 
and \autoref{fig:discstruct_crossec}{\it a}). The difference in irradiation and temperature between the upper and lower side of the disc is 
clearly visible in the surface brightness in Case C orientation, i.e. if the disc bends perpendicular to the line of sight 
(see \autoref{fig:mom1azimut}{\it p}). Interestingly, as we have already seen in the case of the spatially integrated line profile, the projected velocity 
field of edge-on warped discs show negligible deviation from that of an unperturbed disc, which is due to the nature of the projection. 

The surface brightness asymmetry by itself is not necessarily a unique characteristics of warps in circumbinary discs. 
It is known, that in spatially resolved observations of optically thick gas lines the far side of the disc can 
appear to be brighter than the near side even for symmetric discs. However, this is just an observational artifact caused by the projection. 
The surface brightness in optically thick gas line observations depends on the local temperature and the projected emitting area. In flared discs the
emitting are is smaller on the near side of the disc, due to the projection, than on the far side, causing an apparent lower surface brightness on
the near side compared to the far side. Even though the strength of this effect can be comparable to the surface brightness asymmetry caused
by warps in the disc, the kinematics of the disc is only affected by the warp. Therefore kinematical signatures are more robust diagnostics of warps
in spatially resolved gas line observations than the surface-brightness asymmetry alone.

\subsubsection{Effect of inclination on kinematical diagnostics of warps}
As we showed in Section~\ref{sec:line_profiles} and Section~\ref{sec:moment_maps} kinematical signatures are important direct observational diagnostics of warps in circumbinary discs.
However, the strength of the deviation in the projected velocity structure of a warped disc from that of a symmetric Keplerian disc
depends on the inclination both in spatially resolved line profile and in spatially resolved channel maps. The velocity deviation in a warped disc becomes stronger for lower inclinations. This is clearly seen in spatially integrated line profile (see \autoref{fig:specshape}). The reason for that is 
the following.  Let us consider a warped disc in Case A/B orientation, i.e. when the disc bends along the line of sight. The difference in the projected 
velocity of a symmetric ($v_{\rm sym}$) and a warped disc  ($v_{\rm warp}$) along the major axis of the projected disc ellipse depends on the inclination 
angle and the tilt angle such that $v_{\rm sym} - v_{\rm warp} \propto v_{\rm k}(\sin{i} \pm \sin{(i\pm\Delta\beta^\prime)})$, with $v_{\rm k}$ being the
Keplerian velocity. It is easy to show, by expanding the velocity difference in the powers of $i$ and $\Delta\beta^\prime$, that for low inclination the 
velocity difference between a warped and a symmetric disc to linear order is set by $v_{\rm k}\Delta\beta^\prime$ only. In contrast, for high inclination 
angles the velocity difference decreases to linear order as $v_{\rm k}\Delta\beta^\prime(\pi/2-i)$ with the inclination angle. For $i=90^\circ$ the velocity 
difference between an unperturbed and a warped disc becomes $v_{\rm k}(1-\cos{\Delta\beta^\prime})$, i.e. only a few percent of the local Keplerian 
velocity in the strongest warps in our models. 

The orientation of the warp and the inclination of the outer edge of the disc can have an extreme strong effect on the appearance of the disc in 
gas-line observations at low inclination angles as clearly visible in \autoref{fig:channel_maps}. If $\Delta\beta^\prime>i_{\rm out}$, where 
$i_{\rm out}$ denotes the inclination angle at the outer radius of the disc, and the orientation of the warp is such as in Case A
(see \autoref{fig:channel_maps}{\it f,j,n}) the structure of the channel maps of warped discs does not show any peculiarity. In this case the 
presence of the warp can only be inferred from the projected radial velocity curve, i.e. an increasing excess velocity 
towards the inner edge of the disc above that of an unperturbed disc, and the possible surface brightness asymmetry. In contrast, if the disc bends 
in the opposite direction, i.e. the effective inclination angle decreases inwards (Case B), the rotation of the disc seems to change direction at a certain 
radius. The local effective inclination angle of the disc in this case is $i(r) = i_{\rm out} - \beta^\prime(r)$. Thus, at the radius where 
$\beta^\prime(r) = i_{\rm out}$ the projected velocity vanishes, and inwards of the radius, where $\beta^\prime(r) > i_{\rm out}$, the projected velocity 
changes sign, causing the disc to appear as if it would be rotating in the opposite direction than in the outer regions. This apparent change in the
direction of rotation is a characteristic feature of warped disc by which the presence of a warp in the disc is easily identified.

\section{Summary and conclusions}
\label{sec:concl}
 We performed 3D SPH hydrodynamical simulations in combination with 3D radiative transfer calculations to study the observational signatures of 
linear warps in circumbinary discs. The main conclusions of this study can be summarised as follows. 

Our numerical simulations confirm previous analytical results on the dependence of the twist ($\gamma(r)$) and the tilt ($\beta(r)$) angles on the viscosity 
and the initial misalignment between the binary orbital plane and the disc. For a given fixed binary mass ratio and orbital parameters, the tilt angle shows extremely 
weak dependence on viscosity but it depends linearly on the initial misalignment angle. The twist on the other hand is independent of the initial misalignment and
set by the viscosity instead. The observational signatures of warped disc are determined by $\beta^\prime(r)$ and  $\gamma^\prime(r)$ which are the 
equivalent of the tilt and the twist angles in a coordinate system aligned with the plane of the outer edge of the disc.

The thermal structure of warped circumbinary discs shows deviations from that of a symmetric disc both in the vertical and in the azimuthal direction 
due to the non-axisymmetric shadowing of the outer regions by the inner edge of the disc.  The self-shadowing of the disc is characteristically large scale in the
azimuthal direction and results in a non-axisymmetric surface brightness distribution in spatially resolved observations. The surface brightness asymmetry, 
whose azimuthal direction is set by the orientation of the warp, is seen at all inclination angles in the continuum at infrared wavelengths and in the optically 
thick line emission that originates close to the disc surface. In spatially resolved sub-millimetre continuum observations the disc shows a largely symmetric surface
brightness distribution, since the emission originates mostly close to the midplane, where the thermal structure is not significantly affected by the warp.

Warped discs show characteristic deviations in the projected velocity structure from that of a symmetric Keplerian disc. If the disc bends along the line of sight, the effective local inclination angle of the disc decreases or increases depending whether
the disc bends towards or away from the observer. In this case even though the projected velocity map remains symmetric, resembling that of an unperturbed
disc, a deficit or excess emission at high velocities can be seen compared to the velocity pattern of an unperturbed Keplerian disc. In the most extreme case
if the disc bends towards the observer more than the inclination angle of the outer edge of the disc, the inner edge of the disc appears rotating in the opposite
direction than the outer edge of the disc. If the disc bends perpendicular to the line of sight spatially resolved channel maps at the systemic velocity and the zero velocity curve in first moment maps show an S-shaped curvature. In this case the surface brightness asymmetry will appear between the left and right side of the
disc in optically thick lines, causing an asymmetry in the strength of the two Keplerian peaks in the spatially integrated line profile. The strength of the 
kinematical signatures of warps decreases with increasing inclination angle.

Finally, we note that even though in this paper we focused on the specific case of warped {\it circumbinary} discs, the general observational features of warped discs do not depend on the physical mechanism generating the warp. Some details can obviously be different for different tilt and twist structures, but our main conclusions are quite general for any exciting torque, i.e. for any warped disc.

\section*{Acknowledgements}
We are thankful to Giuseppe Lodato, Cathie Clarke and Jim Pringle for fruitful discussions, and Daniel Price for providing the \textsc{phantom} code and useful comments on the manuscript. This work has been supported by the DISCSIM project, grant agreement 341137 funded by the European Research Council under ERC-2013-ADG. \autoref{fig:sph} was produced using \textsc{splash} \citep{2007PASA...24..159P}, a visualisation tool for SPH data. The other figures were generated with the \textsc{python}-based package \textsc{matplotlib} \citep{2007CSE.....9...90H}.


\bibliographystyle{mnras}
\bibliography{ms}



\appendix
\section{Effect of the strength of the warp on the observational signatures}
Here we present synthetic observations calculated from the weakest warp in our hydrodynamic simulations. Since the strength of the perturbation 
decreases with decreasing viscosity ($\alpha$) and decreasing initial misalignment angle ($\beta_0$) the weakest deformation of the disc appears
in our hydrodynamic simulation with  $\alpha=0.02$ and $\beta_0=15^\circ$. 

\begin{figure*}
\center
\includegraphics[width=\textwidth]{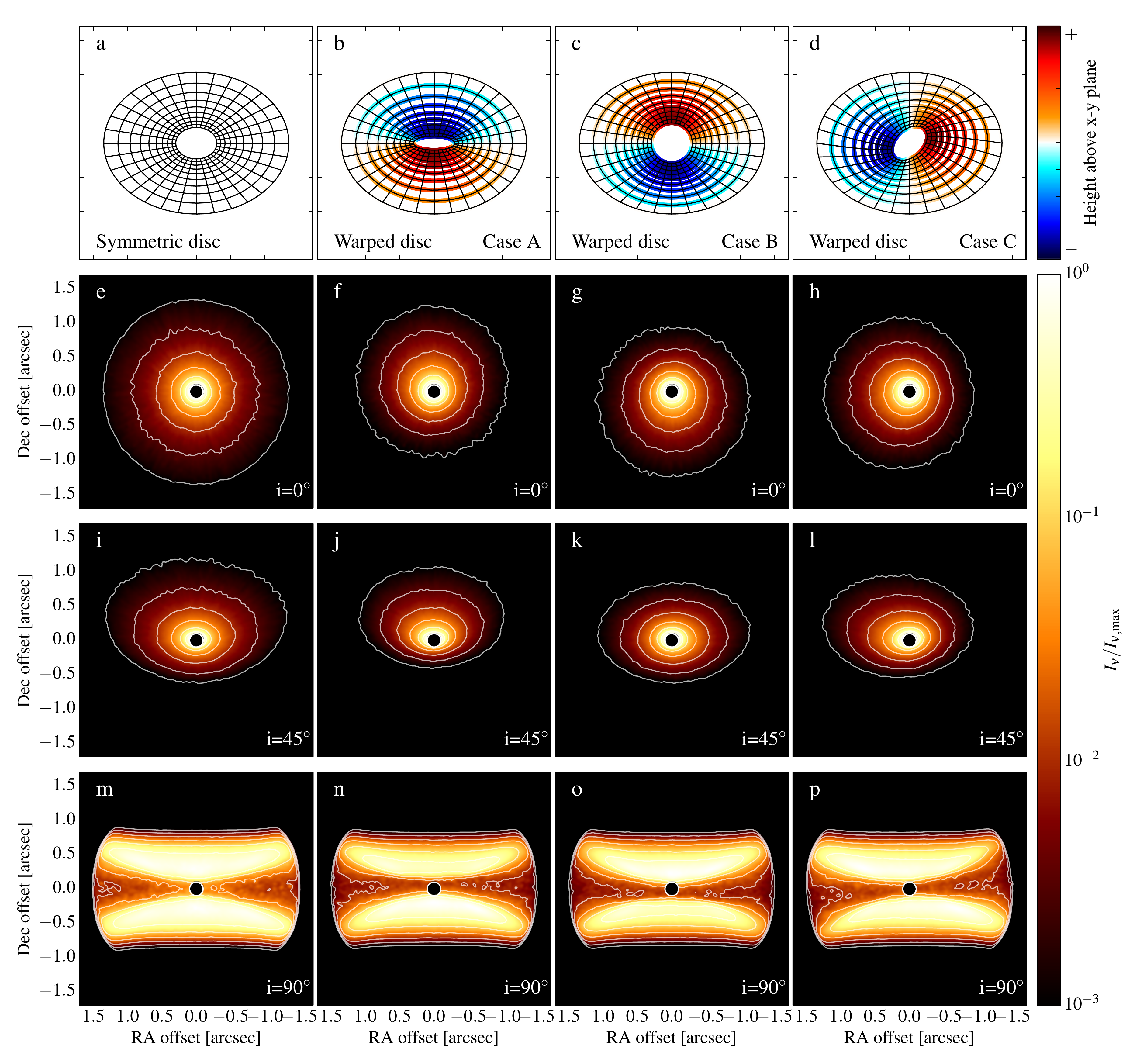}
\caption{Same as \autoref{fig:image_nirscat}, but for the weakest warp in our simulations with $\alpha=0.02$ and $\beta_0=15^\circ$. 
The deviation of the surface brightness distribution of such weakly warped disc from that of that of an unperturbed, symmetric disc is significantly
weaker than in our fiducial model at all inclination angles. The presence of a warp might be inferred from the shift of the stellar position from the 
centre of the disc as seen in scattered light. However, this effect can also be caused by anisotropic scattering if the grains are comparable or larger
than the wavelength of observations. Note that the smaller outer radius of the warped discs compared to the unperturbed model, whose surface density is based on 
our fiducial model (with $\alpha=0.2$ and $\beta_0=30^\circ$) is due to the faster spreading of the disc in the hydrodynamic simulations in the fiducial model. 
}
\label{fig:app_image_nirscat}
\end{figure*}

\begin{figure*}
\center
\includegraphics[width=\textwidth]{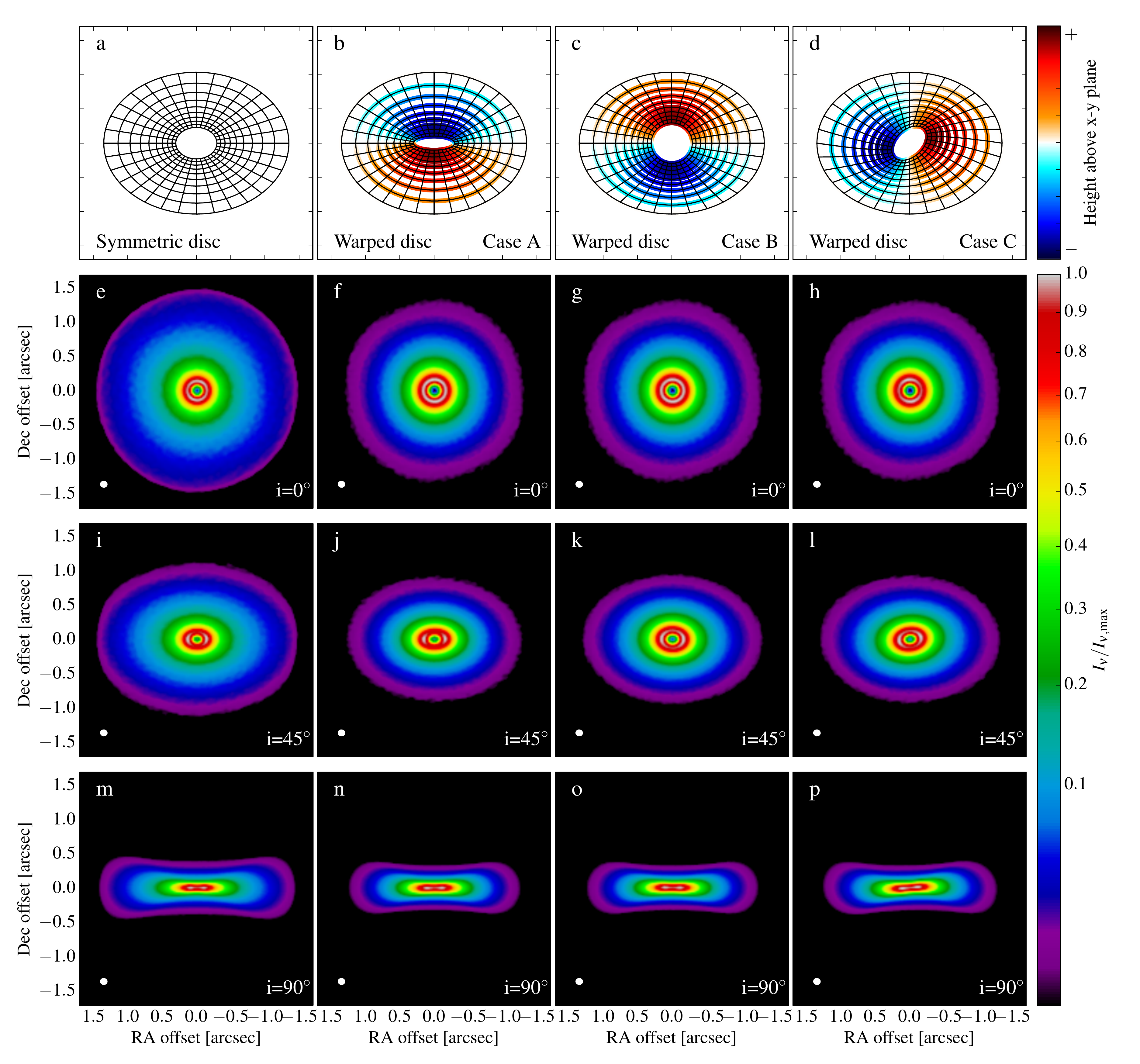}
\caption{Same as \autoref{fig:image_submm_cont}, but for the weakest warp in our simulations with $\alpha=0.02$ and $\beta_0=15^\circ$. 
In most cases such weak warping does not cause any observable deviation in the surface brightness distribution from that of an unperturbed disc.
The warping is only visible in panels {\it l} and {\it p}, i.e, if the disc is bending perpendicular to the line of sight. In this case the warping of the disc
causes the position angle of the isophotes to change with radius. 
}
\label{fig:app_image_submm_cont}
\end{figure*}

\begin{figure*}
\center
\includegraphics[width=\textwidth]{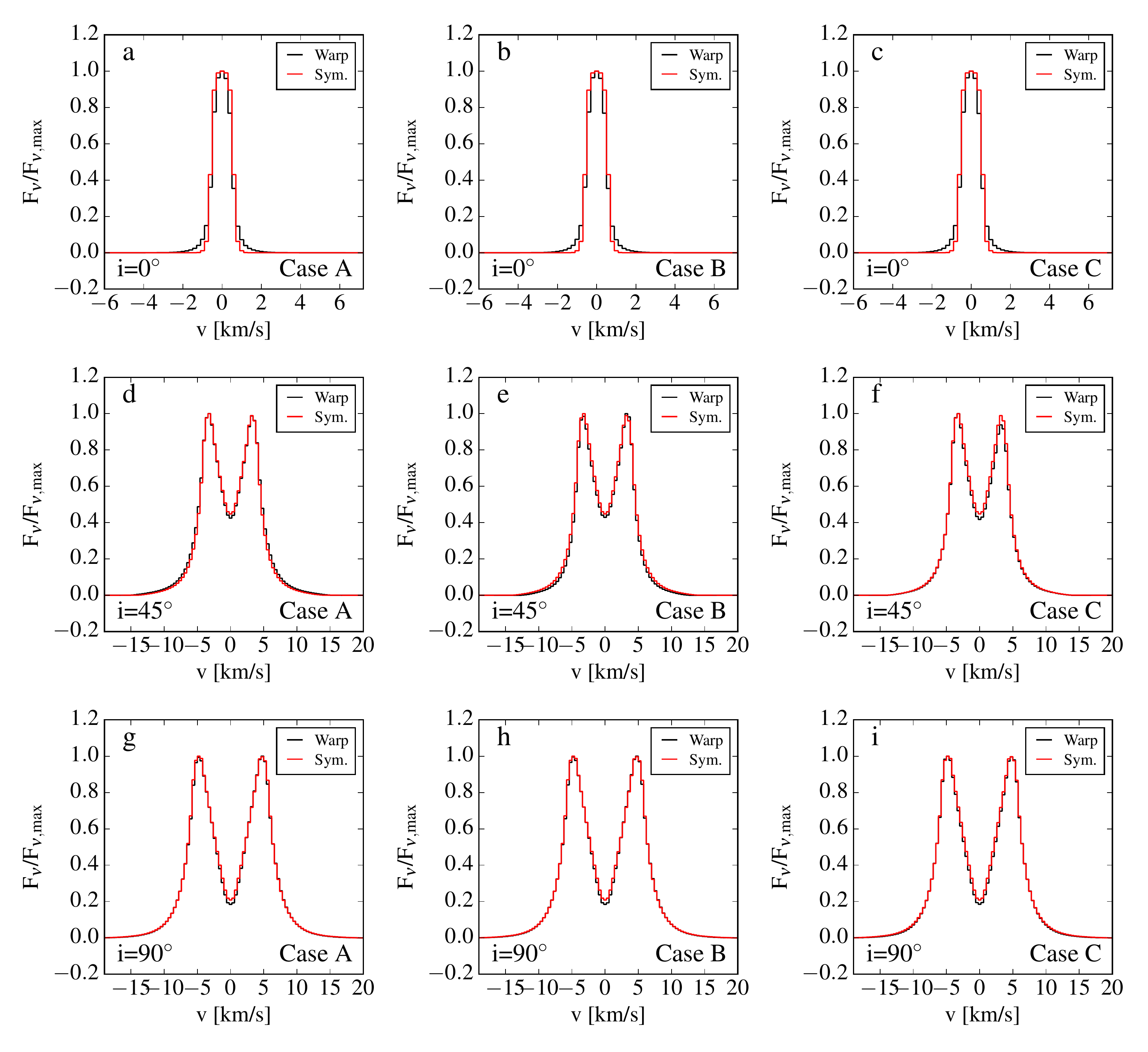}
\caption{Same as \autoref{fig:specshape}, but for the weakest warp in our simulations with $\alpha=0.02$ and $\beta_0=15^\circ$.  
The warping of the disc in this model is so low that the deviation of the line-profile from that of an unperturbed disc is only visible
if the outer disc is seen face-on. At moderate to high inclination angles the line profile of such weakly warped disc and that of an unperturbed
symmetric disc is indistinguishable from each other. In real observations with realistic noise levels the detection of a warp can be
challenging even at very low inclination angles.}
\label{fig:app_specshape}
\end{figure*}

\begin{figure*}
\center
\includegraphics[width=\textwidth]{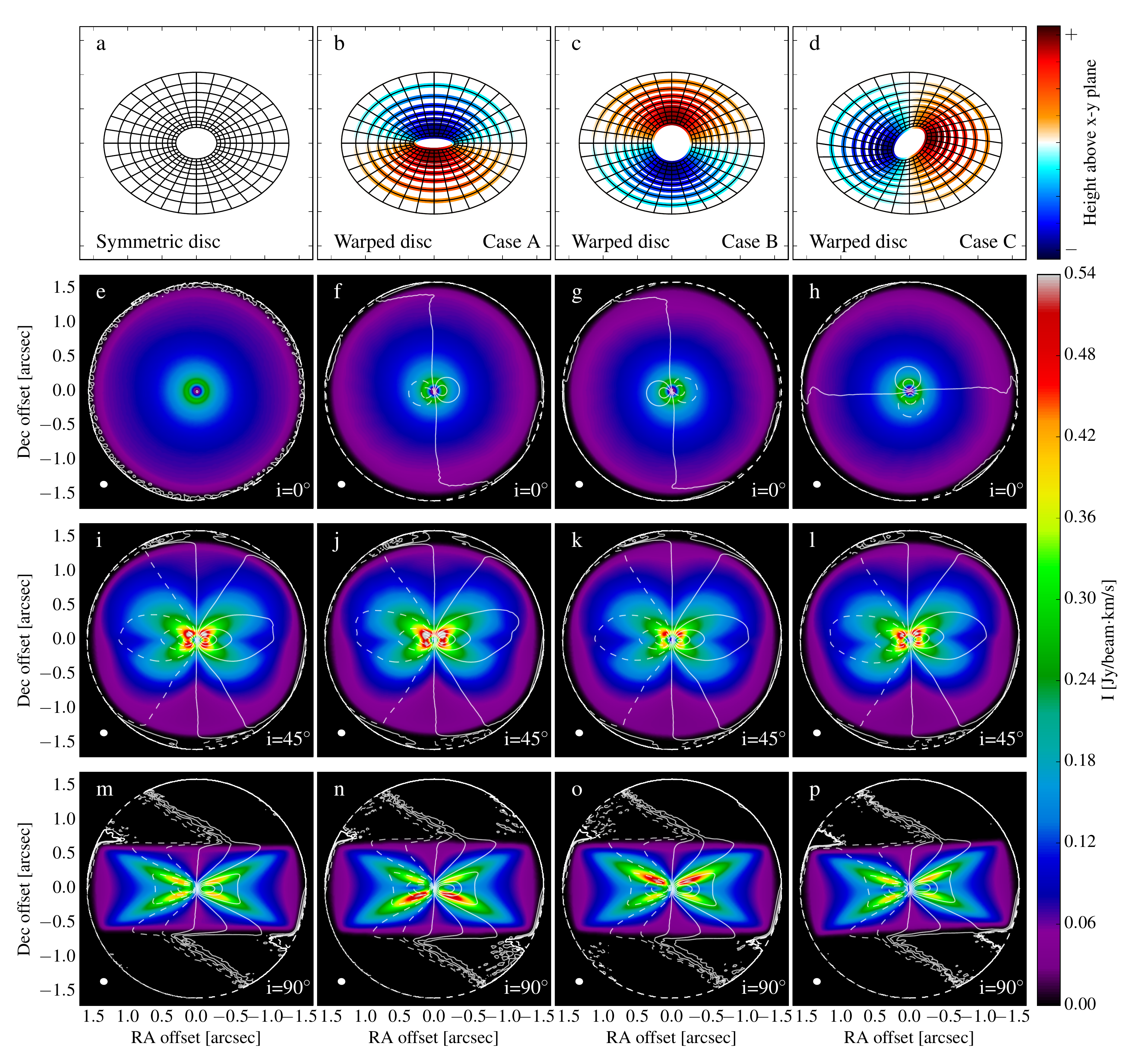}
\caption{Same as \autoref{fig:mom1azimut}, but for the weakest warp in our simulations with $\alpha=0.02$ and $\beta_0=15^\circ$. 
Both the kinematic signatures and the surface brightness asymmetry are significantly weaker than in the fiducial model with $\alpha=0.2$ and $\beta_0=30^\circ$.
The surface brigthness asymmetry is only visible clearly when the disc is seen at a very high inclination angle and the disc is bending along the 
line of sight (Panel {\it n, o}). The strength of the kinematic signatures decreases with increasing inclination angle. 
}
\label{fig:app_mom1azimut}
\end{figure*}

\begin{figure*}
\center
\includegraphics[width=\textwidth]{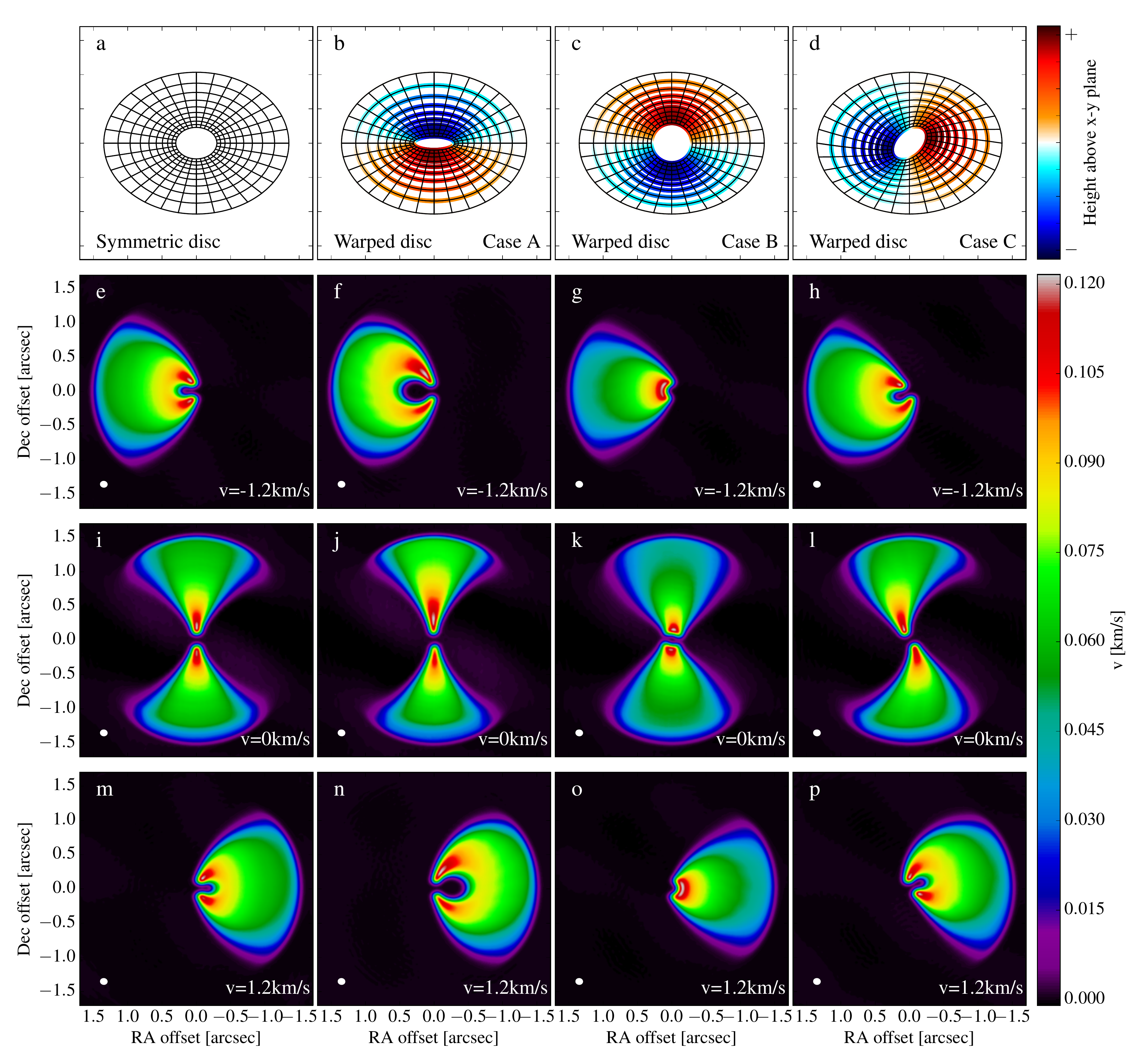}
\caption{Same as \autoref{fig:channel_maps}, but for the weakest warp in our simulations with $\alpha=0.02$ and $\beta_0=15^\circ$. 
In this case $\beta^\prime(r_{\rm in})=5.9^\circ$. Since the assumed inclination angle is larger ($i=10^\circ$) no apparent counter rotation occurs.
However, the warping of the disc is visible if the disc bends perpendicular to the line of sight (Case C) as the emission at the zero velocity channel
becomes twisted. If the disc bends along the line of sight (Case A/B), warping of the disc becomes difficult to recognise, especially in Case A.  
 }
\label{fig:app_channel_maps}
\end{figure*}

\bsp	
\label{lastpage}
\end{document}